\documentclass[pre,aps,twocolumn,showpacs]{revtex4}
\usepackage{epsfig,latexsym}
\usepackage[hypertex]{hyperref}
\begin{document}

\title{Crossing intervals of non-Markovian Gaussian processes}

\author{Cl\'ement Sire}
\email{clement.sire@irsamc.ups-tlse.fr}
\affiliation{Laboratoire de Physique Th\'eorique -- IRSAMC, CNRS,
Universit\'e Paul Sabatier, 31062 Toulouse, France}

\begin{abstract}
We review the properties of time intervals between the crossings at
a level $M$ of a smooth stationary Gaussian temporal signal. The
distribution of these intervals and the persistence are derived
within the Independent Interval Approximation (IIA). These results
grant access to the distribution of extrema of a general Gaussian
process. Exact results are obtained for the persistence exponents
and the crossing interval distributions, in the limit of large
$|M|$. In addition, the small time behavior of the interval
distributions and the persistence is calculated analytically, for
any $M$. The IIA is found to reproduce most of these exact results
and its accuracy is also illustrated by extensive numerical
simulations applied to non-Markovian Gaussian processes appearing in
various physical contexts.
\end{abstract}


\maketitle
\section{Introduction}

The persistence of a temporal signal $X$ is the probability that
$X(t')$ remains below (or above) a given level $M$, for all times
$t'\in[0,t]$. This problem has elicited a large body of work by
mathematicians \cite{adler,BL,rice,kean,wong,slepian,theta2,merca}, and physicists,
both theorists
\cite{SM,AB1,BD1,per1,per2,per3,iia,globalc,global,red2,krug,csp,red1}
and experimentalists \cite{breath,lq,soap,diff1d,surf}. Persistence
properties have been measured in as different systems as breath
figures \cite{breath}, liquid crystals \cite{lq}, laser-polarized
\textit{Xe} gas \cite{diff1d}, fluctuating steps on a \textit{Si} surface
\cite{surf}, or soap bubbles \cite{soap}.

Although the persistence is a very natural and easy quantity to
define -- in a sense, its physical meaning is easier to explain to
the layman than the meaning of a two-point correlation function --
this quantity turns out, in practice, to be extremely complex to
deal with analytically. In fact, exact results have been obtained in
a very limited number of cases, as far as non-Markovian processes
are concerned \cite{kean,wong,theta2}.

The mathematical literature has mainly focused on evaluating the
persistence
\begin{equation}
P_<(t)={\rm Prob}(X(t')<M,\quad t'\in[0,t]),
\end{equation}
mostly for Gaussian processes (\emph{i.e.} processes $X(t)$ for
which the joint distribution of $X(t_1),..., X(t_n)$ is Gaussian)
and for large $|M|$, a regime where efficient bounds or equivalent
have been obtained \cite{adler,BL,slepian}. Recently \cite{merca},
and for Gaussian processes only, a numerical method to obtain
valuable numerical bounds has been extended to all values of $M$,
although the required numerical effort can become quite considerable
for large $t$.

Physicists have also concentrated their attention to Gaussian
processes \cite{per1,per2,per3,iia,red2,csp}, which are often a good
or {\it exact} description of actual physical processes. For
instance, the total magnetization in a spin system
\cite{globalc,global}, the density field of diffusing particles
\cite{iia}, or the height profile of certain fluctuating interfaces
\cite{krug,diff1d,surf} are true temporal Gaussian processes. Two
general methods have been developed, focusing on the case $M=0$,
which applies to many physical situations. The first one
\cite{per1,per2,per3} is a perturbation of the considered process
around the Markovian Gaussian process, which has been extended to
small values of $|M|$ \cite{per2}. Within this method, only the
large time asymptotics of $P_<(t)$ is known, leading to the
definition of the persistence exponent (see below). The alternative
method, using the Independent Interval Approximation (IIA)
\cite{iia,csp}, gives very accurate results for ``smooth''
processes, that is, processes having a continuous velocity.
Initially, the IIA remained restricted to the case $M=0$ \cite{iia},
but it has been recently generalized to an arbitrary level $M$
\cite{csp}.

The persistence probability is also intimately related to another
important physical quantity: the probability distribution of the
extrema of the considered process $X$. For instance, the quantity
$P_<(t)$ can be also viewed as the probability that the
\emph{maximum} of $X$ in the interval $[0,t]$ has remained below the
level $M$. Thus, the distribution of the maximum of $X$ is simply
the derivative with respect to $M$ of $P_<(t)$. The distribution of
the extrema has been analytically obtained for the Brownian process
\cite{max1,max2}, but its derivation remains a formidable task for
general non-Markovian Gaussian processes. On this account, the
persistence problem is also related to extreme value statistics
\cite{Gumbel,Galambos}, which is a notoriously difficult problem for
correlated variables, and which has recently attracted a lot of
attention among physicists \cite{extreme}.

Hence, the persistence problem has obvious applications in many
other applied and experimental sciences, where one has to deal with
data analysis of complex statistical signals. For instance,
statistical bounds of noisy signals are extremely useful for image
processing (for instance in medical imaging or astrophysics
\cite{image}), in order to obtain cleaner images by correcting
spurious bright or dark pixels \cite{adler,merca}. In general, it is
important to be able to evaluate the maximum of a correlated
temporal or spatial signal originating from experimental noise. The
same question can arise when the signal lives in a more abstract
space. For instance, in the context of genetic cartography,
statistical methods to evaluate the maximum of a complex signal has
been exploited to identify putative quantitative trait loci
\cite{lander}. Finally, this same problem arises in econophysics or
finance, where the probability for a generally strongly correlated
financial signal to remain below or above a certain level is clearly
of great interest \cite{boupot}.

In the present paper, we are interested in the persistence of a
non-Markovian Gaussian process, which can be either stationary or
scale-invariant. More specifically, this study focuses on the
properties of the distributions of time intervals during which the
considered process remains below or above a given level $M$. We
shall see that these distributions are simply related to the
persistence itself, and contain valuable physical information.

We now summarize the content of the present work, and mention
briefly our main results. In section~\ref{II}, we introduce our main
quantities of interest, and among them: \emph{i.\,}the persistence
$P_>(t)$ and $P_<(t)$ which measure the probability to remain above
or below the level $M$ up to time $t$; \emph{ii.\,}the associated
distributions of $\pm$ time intervals, $P_\pm(t)$, during which the
process remains above or below the level $M$. For a general level
$M$, we then shortly review the IIA calculation introduced in
\cite{csp}, which leads to analytical approximate expressions for
the Laplace transform of $P_\pm(t)$, $P_>(t)$, and $P_<(t)$. In
fact, we present a simpler formulation compared to \cite{csp}, as
well as new additional results, which permit a fast and efficient
numerical implementation of the IIA results. In section~\ref{III},
we first introduce several examples of physically relevant smooth
Gaussian processes, on which the general results of this paper are
numerically tested. Within the IIA, we then present the calculation
of the persistence exponents $\theta_\pm(M)$, defined, for
stationary processes, by $P_\pm(t)\sim\exp(-\theta_\pm\,t)$, when
$t\to +\infty$. We also obtain exact estimates for $\theta_\pm(M)$,
for large $M>0$, including exact bounds for $P_>(t)$ and
$\theta_+(M)$, and compute the exact asymptotic distributions
$P_\pm(t)$. For large $M$, $P_+(t)$ takes the form of the Wigner
distribution, whereas $P_-(t)$ becomes Poissonian. All these results
are correctly reproduced by the IIA, except for the large $M$
asymptotics of $\theta_+(M)$. Finally, in section~\ref{secdist}, we
obtain the exact small time behavior of $P_\pm(t)$, $P_>(t)$, and
$P_<(t)$, for ``very smooth'' processes (a term to be defined in the
next section~\ref{II}), and find that the IIA again reproduces these
exact results. For marginally smooth processes, we also obtain exact
results, but the IIA is not exact in this case, although its results
remain qualitatively correct and even quantitatively accurate. Many
of these results are also obtained by means of simple heuristic
arguments, helping to understand them on physical grounds. This
study is also illustrated by means of extensive numerical
simulations, revealing a very satisfying accuracy of the IIA for
moderate $M$, in a regime where, contrary to the limit
$M\to\pm\infty$, there are no available exact results for
$P_\pm(t)$, $P_>(t)$, and $P_<(t)$.

\section{Independent Interval Approximation (IIA)}\label{II}

In this section, we introduce our main physical quantities of
interest -- interval distributions, persistence, sign
autocorrelation, constrained densities of crossings -- and relate
their general properties. We then summarize the IIA calculation
introduced in \cite{csp}, and obtain more explicit expressions for
the interval distributions and the persistence probability, in the
case of a Gaussian process. These new results will prove useful in
the next sections more specifically devoted to the interval
distributions.

\subsection{Introductory material and notations}
One considers a stationary non-Markovian Gaussian process $X(t)$ of
zero mean and unit variance. Its distribution at any time is then
\begin{equation}
g(X)=\frac{{\rm e}^{-\frac{X^2}{2}}}{\sqrt{2\pi}},\label{gauss1}
\end{equation}
and we define its cumulative distribution as,
\begin{equation}
G(X)=\int_{-\infty}^Xg(x)\,dx=1-\bar G(X).\label{gauss2}
\end{equation}

Due to its Gaussian nature, such a stationary process is entirely
characterized by its two-point correlation function,
\begin{equation}
f(t)=\langle X(t+t')\,X(t')\rangle.
\end{equation}
It is understood that the process starts from $t=-\infty$, so that
the distribution of the initial condition $X(0)$ at $t=0$ is given
by Eq.~(\ref{gauss1}). In addition, all derivatives of $X(0)$, when
they exists, are random Gaussian variables of zero average and
second moment
\begin{equation}
\left\langle \left[X^{(n)}(0)\right]^2\right\rangle=(-1)^n f^{(2n)}(0),
\end{equation}
where the superscript $(n)$ refers to a derivative of order $n$.

The process is assumed to be ``smooth'', although this constraint
will be sometimes relaxed in the present paper. By ``smooth'', we
mean that the velocity of the process is a continuous function of
time. ``Very smooth'' processes will have a differentiable velocity.
In particular, the most general stationary \emph{Markovian} Gaussian
process defined by the equation of motion,
\begin{equation}
\frac{dX}{dt}=-\lambda X+\sqrt{2\lambda}\,\eta,\label{markov}
\end{equation}
does not belong to this class of smooth processes ($\eta(t)$ is a
Gaussian $\delta$-correlated white noise). In practice, a process is
smooth if its correlator $f(t)$ is twice differentiable at $t=0$,
and is very smooth if the fourth derivative of $f(t)$ exists at
$t=0$. The process introduced in Eq.~(\ref{markov}) has a correlator
\begin{equation}
f(t)=\exp(-\lambda |t|),\label{markovcor}
\end{equation}
which has a cusp at $t=0$, and is thus not even differentiable at $t=0$.

Throughout this paper, we will be interested in the probability that
the process $X$ remains below or above a certain threshold $M$, for
all times in the interval $[0,t]$. In particular, the smoothness of
the process ensures that the $M$-crossings (\textit{i.e.} the times
for which $X(t)=M$) are well separated, with a finite mean
separation between them, denoted by $\tau$. For the Markovian
process mentioned above, the $M$-crossings are distributed on a
fractal set of dimension $1/2$, and the mean interval is then
$\tau=0$. For smooth process, $\tau$ can be computed by evaluating
the mean number of $M$-crossings during a time interval of length
$t$,
\begin{equation}
N(t)=\left\langle
\int_0^t|X'(t')|\delta(X(t')-M)\,dt'\right\rangle=\frac{t}{\tau}.
\label{Nt}
\end{equation}
The correlation functions between the position $X$ and the velocity
$X'$ is
\begin{equation}
\langle X(t)\,X'(t')\rangle=-f'(t-t').
\end{equation}
By time-reversal symmetry, and since $f(t)$ is twice differentiable
at $t=0$, one has
\begin{equation}
\langle X(t)\,X'(t)\rangle=-f'(0)=0,
\end{equation}
so that the position and the velocity are uncorrelated at equal
time. The distribution of the velocity $X'(t)$ is a Gaussian of zero
mean and second moment
\begin{equation}
\langle X'^2(t)\rangle=-f''(0)=a_2,
\end{equation}
so that
\begin{equation}
\langle |X'(t)|\rangle=\sqrt{\frac{2a_2}{\pi}}.
\end{equation}
Finally, $N(t)=\langle |X'(t)|\rangle g(M){\times} t$, which leads to
\begin{equation}
\tau=\frac{\pi}{\sqrt{a_2}}\,{\rm e}^{\frac{M^2}{2}}.\label{taugauss}
\end{equation}

We also define the distributions of time intervals between
$M$-crossings, $P_+(t)$ and $P_-(t)$, during which the process
remains respectively above and below the level $M$. The means of
these two kinds of intervals are defined by
$\tau_\pm=\int_0^{+\infty}t\, P_\pm(t)\,dt$, and are related to
$\tau$. Indeed, since there are as many $+$ and $-$ intervals, as
they simply alternate, one has
\begin{equation}
\tau=\frac{\tau_++\tau_-}{2}.
\end{equation}
In addition, $\tau_+/\tau_-$ is also equal to the ratio of the times
spent by the process $X$ above and below the level $M$,
\textit{i.e.}, $\bar G(M)/G(M)$. Finally, we obtain
\begin{eqnarray}
\tau_-=2\tau\, G(M),\quad\tau_+=2\tau \,\bar G(M).\label{t2}
\end{eqnarray}

We now introduce the persistence of the process $X$, defined as the
probability that it does not cross the level $M$ during the time
interval $[0,t]$. More precisely, we define $P_>(t)$ and $P_<(t)$ as
the persistence, knowing that the process started respectively above
and below the level $M$. In other word, $P_>(t)$ is also the
probability that the process remains above the level $M$ during the
considered time interval, and $P_<(t)$ is the probability that $X$
remains below the threshold $M$. Note that the persistence probes
the entire history of the process, and is therefore a
\emph{infinite-points} correlation function: for instance $P_<(t)$
is the probability that the process remains below the level $M$
between the times $0$ and $dt$, \emph{and} $dt$ and $2\,dt$,...
\emph{and} $t-dt$ and $t$. For non-Markovian processes, it is thus
understandable that this quantity is extremely difficult to treat
analytically, and there are very few examples where $P_>(t)$ or
$P_<(t)$ can be actually computed \cite{BL,merca,theta2}.

$P_>(t)$ and $P_<(t)$ are intimately related to the interval
distributions $P_+(t)$ and $P_-(t)$ by the relations \cite{csp},
\begin{eqnarray}
P_>(t)&=&\tau_+^{-1}\int_t^{+\infty}(t'-t)P_+(t')\,dt',\label{pnew1}\\
P_<(t)&=&\tau_-^{-1}\int_t^{+\infty}(t'-t)P_-(t')\,dt'.\label{pnew2}
\end{eqnarray}
Indeed, if $X(t)$ has remained below the level $M$ up to time $t$,
it belongs to a $-$ interval of duration $t'>t$, starting at an
initial position uniformly distributed between $0$ and $t'-t$, an
interpretation which leads to Eq.~(\ref{pnew2}). The above
expressions can also be differentiated twice, giving
\begin{eqnarray}
P_+(t) &=& \tau_+\, P_>''(t), \label{pnew3}\\
P_-(t) &=& \tau_- \,P_<''(t).  \label{pnew4}
\end{eqnarray}
If $M=0$ \cite{iia}, by symmetry of the process under the
transformation $X\to -X$, one has $P_>(t)=P_<(t)$ and
$P_+(t)=P_-(t)$. In general, we have the symmetrical relations,
\begin{eqnarray}
P_>(t,M)=P_<(t,-M),\quad P_+(t,M)=P_-(t,-M),
\end{eqnarray}
so that we will restrain ourselves to the case $M\geq 0$.

The knowledge of $P_>(t)$ and $P_<(t)$ also provides valuable
information on the distribution of the extrema of the considered
process. The statistical properties of extremal events is an active
field of research among mathematicians \cite{Gumbel,Galambos} and
physicists \cite{extreme}. By definition, $P_<(t)$ is also the
probability that the maximum of the process in a time interval of
duration $t$ is less than $M$, and $P_>(t)$ is the probability that
the minimum of the process remains bigger than $M$ for all times in
$[0,t]$. Hence, defining $P_{\rm max}(M,t)$ and $P_{\rm min}(M,t)$
as the distribution of the maximum and minimum of the considered
process, one has,
\begin{eqnarray}
P_{\rm min}(M,t) &=&-\frac{\partial}{\partial M} P_>(t),  \label{pmin}\\
P_{\rm max}(M,t) &=&\frac{\partial}{\partial M} P_<(t).  \label{pmax}
\end{eqnarray}
Note finally that
Eqs.~(\ref{pnew1},\ref{pnew2},\ref{pnew3},\ref{pnew4},\ref{pmin},\ref{pmax})
are in fact valid for \emph{any stationary process}, not necessarily
Gaussian.

Before presenting an approximate method leading to analytical
expressions for the different quantities introduced above, we need
to define two quantities which will prove useful in the following.
We start with  the autocorrelation function of $\theta [M-X(t)]$
($\theta$ is Heaviside's function: $\theta(x)=1$, if $x>0$;
$\theta(x)=0$, if $x<0$; $\theta(0)=1/2$),
\begin{eqnarray}
A_>(t)&=&\langle \theta [X(t)-M]\, \theta [X(0)-M]\rangle,\label{ap}\\
A_<(t)&=&\langle \theta [M-X(t)]\, \theta [M-X(0)]\rangle,\nonumber\\
      &=& 2G(M)-1+A_>(t), \label{am}
\end{eqnarray}
where the last relation is obtained by using
$\theta(x)=1-\theta(-x)$. In addition, since the process is
invariant under the transformation $X\to -X$, one also has,
\begin{equation}
A_>(M,t)=A_<(-M,t).
\end{equation}
For a Gaussian process, these quantities can be explicitly expressed
in terms of the correlation function $f(t)$ \cite{potts},
\begin{eqnarray}
A_>(t)&=&  \int_{M}^{+\infty} g(x)\,\bar G\left(\frac{M-x
f(t)}{\sqrt{1-f^2(t)}}\right)\,dx,\label{apf}\\
A_<(t)&=&  \int_{-\infty}^{M} g(x)\,G\left(\frac{M-x
f(t)}{\sqrt{1-f^2(t)}}\right)\,dx.\label{amf}
\end{eqnarray}
For $M=0$, these integrals can be explicitly performed \cite{iia},
giving
\begin{equation}
A_>(t)=A_<(t)=\frac{1}{4}+\frac{1}{2\pi}\arcsin(f(t)).
\end{equation}
Finally, the time derivative of $A_>(t)$ and $A_<(t)$ can be simply
written as
\begin{equation}
A_>'(t)=A_<'(t)= \frac{1}{2\pi}\frac{f'(t)}{\sqrt{1-f^2(t)}}
\exp\left(-\frac{M^2}{1+f(t)}\right). \label{dapf}
\end{equation}

We now introduce $N_>(t)$ and $N_<(t)$, the mean number of
$M$-crossings in the time interval $[0,t]$, knowing that the process
started from $X(0)>M$ and $X(0)<M$, respectively. These quantities
satisfy the sum rule,
\begin{equation}
G(M)\,N_<(t)+\bar G(M)\,N_>(t)=N(t)=\frac{t}{\tau},\label{sumrule}
\end{equation}
which expresses the fact that a $M$-crossing is either crossed from
above, or from below $M$. Again, it is clear that
\begin{equation}
N_>(M,t)=N_<(-M,t).
\end{equation}
In addition, for large time $t$,
\begin{equation}
N_>(t)\sim N_<(t)\sim N(t)=\frac{t}{\tau}, \quad t\to+\infty,\label{as}
\end{equation}
since the initial position of $X(t)$ becomes irrelevant when
$t\to+\infty$. On the other hand, for short time,
\begin{equation}
N_>(t)\sim\frac{t}{\tau_+},\quad
N_<(t)\sim\frac{t}{\tau_-}, \quad t\to 0,\label{smalln}
\end{equation}
which expresses the fact that the probability per unit time to meet
the first $M$-crossing is respectively $\tau_+^{-1}$ and
$\tau_-^{-1}$, for $+$ and $-$ intervals. Note that both asymptotics of
Eqs.~(\ref{as},\ref{smalln}) are consistent with the sum rule of
Eq.~(\ref{sumrule}). For a Gaussian process, $N_>(t)$ and $N_<(t)$
can be calculated after introducing the correlation matrix of the
Gaussian vector $(X(t),X(0),X'(t))$, which reads
\begin{equation}
{\cal C}(t)=\left(\begin{array}{ccc}
   1 & f(t) & 0 \\
   f(t) & 1 & f'(t) \\
   0 & f'(t) & -f''(0)
\end{array}\right).
\end{equation}
For instance, in the same spirit as Eq.~(\ref{Nt}), one has
\begin{equation}
N_<(t)=G^{-1}(M)\int_{0}^{t}\langle |X'(t')| \rangle_{<}\,dt',\label{Nm}
\end{equation}
where $\langle |X'(t)| \rangle_{<}$ is the average of the velocity
modulus, knowing that $X(t)=M$, and averaged over $X(0)<M$:
\begin{equation}
\langle |X'(t)| \rangle_{<}=\int_{-\infty}^{M}dx_0
\int_{-\infty}^{+\infty}dv \,\frac{|v|{\rm e}^{-\frac{1}{2}{\bf
U}^\dag{\cal C}^{-1}{\bf U}}}{(2\pi)^{3/2}\sqrt{\det {\cal C}}},
\end{equation}
where ${\bf U}=(M,x_0,v)$. $N_>(t)$ is similarly defined as
\begin{equation}
N_>(t)=\bar G^{-1}(M)\int_{0}^{t}\langle |X'(t')| \rangle_{>}\,dt',\label{Np}
\end{equation}
with
\begin{eqnarray}
\langle |X'(t)| \rangle_{>}&=&\int_{M}^{+\infty}dx_0
\int_{-\infty}^{+\infty}dv \,\frac{|v|{\rm e}^{-\frac{1}{2}{\bf
U}^\dag{\cal C}^{-1}{\bf U}}}{(2\pi)^{3/2}\sqrt{\det {\cal C}}}\\
&=&\tau^{-1}-\langle |X'(t)| \rangle_{<}.\label{sumX}
\end{eqnarray}
$\langle |X'(t)| \rangle_{<}$ can be written more explicitly,
\begin{eqnarray}
\tau\,\langle |X'(t)| \rangle_{<}=G(b)+a
\left[\frac 12-G\left(a\,b\right)\right]
{\rm e}^{-\frac{M^2}{2}.\frac{1-f}{1+f}},\label{Xm}
\end{eqnarray}
with
\begin{eqnarray}
a(t)&=&\frac{|f'(t)|}{\sqrt{a_2\left(1-f^2(t)\right)}}, \label{defa}\\
b(t)&=&M\frac{1-f(t)}{\sqrt{1-f^2(t)-f'^2(t)/a_2}}, \label{defb}
\end{eqnarray}
where $a_2=-f''(0)$. Using Eq.~(\ref{sumX}), one finds a similar
expression for $\langle |X'(t)| \rangle_{>}$,
\begin{eqnarray}
\tau\,\langle |X'(t)| \rangle_{>}=\bar G(b)-a
\left[\frac 12-G\left(a\,b\right)\right]
{\rm e}^{-\frac{M^2}{2}.\frac{1-f}{1+f}}.\label{Xp}
\end{eqnarray}
When $t\to +\infty$, $a(t)\to 0$ and $b(t)\to M$, so
that \footnote{Note that there was a typographical error in
\cite{csp}, and that one should replace $\langle |X'(t)|
\rangle_{<}$ by $\frac{\langle |X'(t)| \rangle_{<}}{G(M)}$ in
Eqs.~(28,29) of \cite{csp}.}
\begin{equation}
\langle |X'(t)| \rangle_{<}\sim  \frac{G(M)}{\tau},\quad
\langle |X'(t)| \rangle_{>}\sim \frac{\bar G(M)}{\tau}.
\end{equation}
Using Eqs.~(\ref{Nm},\ref{Np}), one recovers the asymptotics of
Eq.~(\ref{as}). On the other hand, when $t\to 0$, we have $a(t)\to
1$ and $f(t)\to 1$, which leads to
\begin{equation}
\langle |X'(t)| \rangle_{<}\sim \langle |X'(t)| \rangle_{>}\sim
\frac{1}{2\tau},
\end{equation}
and we recover Eq.~(\ref{smalln}).

\subsection{Derivation of the IIA distributions}

In the previous subsection, we have introduced the so far unknown
interval distributions $P_+(t)$ and $P_-(t)$ which are intimately
related to the persistence probabilities $P_>(t)$ and $P_<(t)$
(through Eqs.~(\ref{pnew1},\ref{pnew2},\ref{pnew3},\ref{pnew4})) and
the distribution of extrema of the process (see
Eqs.~(\ref{pmin},\ref{pmax})). On the other hand, for a Gaussian
process, we have computed explicitly the autocorrelation $A_>(t)$
(and $A_<(t)$) and the constrained number of $M$-crossings $N_>(t)$
(and $N_<(t)$) as a function of the correlator $f(t)$ (see
Eqs.~(\ref{dapf},\ref{Xm},\ref{Xp})). We will now try to relate the
two unknown interval distributions to these two known quantities.

Let us define $P_<(N,t)$ and $P_>(N,t)$, as the probability that
there are exactly $N$ $M$-crossings in the interval $[0,t]$,
starting respectively from $X(0)<M$ and $X(0)>M$. By definition of
$A_>(t)$, one has
\begin{eqnarray}
A_>(t)&=& G(M)\sum_{n=0}^{+\infty}P_>(2n,t),\label{conda1}\\
A_<(t)&=& \bar G(M)\sum_{n=0}^{+\infty}P_<(2n,t),\label{conda2}
\end{eqnarray}
since $X(t)$ is on the same side of $M$ as $X(0)$ if and only if the
number of $M$-crossings in the interval $[0,t]$ is even. $N_>(t)$
and $N_<(t)$ can also be simply written as
\begin{eqnarray}
N_>(t)&=&\sum_{n=0}^{+\infty}n\,P_>(n,t),\label{condn1}\\
N_<(t)&=&\sum_{n=0}^{+\infty}n\,P_<(n,t).\label{condn2}
\end{eqnarray}
Note that, by definition, one has $P_>(0,t)=P_>(t)$ and
$P_<(0,t)=P_<(t)$.

Our central approximation now consists in assuming that the {\it
interval lengths between $M$-crossings are uncorrelated}
\cite{iia,csp}. This will lead to closed relation between
($P_<(N,t)$, $P_>(N,t)$) and $P_\pm(t)$. Using
Eqs.~(\ref{conda1},\ref{conda2},\ref{condn1},\ref{condn2}), we will
then obtain an explicit expression of $P_\pm(t)$ as a function of
$A_>(t)$ (or $A_<(t)$) and $N_>(t)$ (or $N_<(t)$).

Let us consider an odd value of $N=2n-1$ $(n\geq 1)$. Using the IIA,
we obtain
\begin{eqnarray}
&P_<(2n-1,t)={\tau_-^{-1}}\int_0^tdt_1\,Q_-(t_1){\times}
\nonumber\\
&\int_{t_1}^t dt_2\,P_+(t_2-t_1) \int_{t_2}^t
dt_3\,P_-(t_3-t_2)\cdots \nonumber\\
&\int_{t_{2n-3}}^t dt_{2n-2}\,P_+(t_{2n-2}-t_{2n-3}){\times}
\nonumber\\
&\int_{t_{2n-2}}^t
dt_{2n-1}\,P_-(t_{2n-1}-t_{2n-2})Q_+(t-t_{2n-1}),\label{conv1}
\end{eqnarray}
where
\begin{equation}
Q_\pm(t)=\int_t^{+\infty} P_\pm(t')\,dt',\label{defq}
\end{equation}
is the probability that a $\pm$ interval is larger than $t$.
Eqs.~(\ref{conv1}) expresses the fact that to find $2n-1$ crossings
between $0$ and $t$ starting from $X(0)<M$, we should find a first
crossing at $t_1$ (and hence an initial $-$ interval of length
bigger than $t_1$, with probability $\tau_-^{-1}Q_-(t_1)$), followed
by a $+$ interval of length $t_2-t_1$, and so on, up to a last
crossing time $t_{2n-1}$, associated to a $-$ interval of length
$t_{2n-1}-t_{2n-2}$. Finally, there should not be any further
crossing between $t_{2n-1}$ and $t$, hence the last factor
$Q_+(t-t_{2n-1})$. All these events have been treated as
independent, so that there probabilities simply factor: this is the
core assumption of the IIA. For even $N=2n$ $(n\geq 1)$, one obtains
a similar expression,
\begin{eqnarray}
&P_<(2n,t)={\tau_-^{-1}}\int_0^t dt_1\,Q_-(t_1){\times}
\nonumber\\
&\int_{t_1}^t dt_2\,P_+(t_2-t_1) \int_{t_2}^t
dt_3\,P_-(t_3-t_2)\cdots \nonumber\\
&\int_{t_{2n-2}}^t dt_{2n-1}\,P_-(t_{2n-1}-t_{2n-2}){\times}
\nonumber\\
&\int_{t_{2n-1}}^t
dt_{2n}\,P_+(t_{2n}-t_{2n-1})Q_-(t-t_{2n}).\label{conv2}
\end{eqnarray}

For a given function of time $F(t)$, one defines its Laplace
transform, $\hat F(s)=\int_0^{+\infty}F(t)\,{\rm e}^{-st}\,dt$. The
convolution products in Eqs.~(\ref{conv1},\ref{conv2}) take a much
simpler form in the Laplace variable $s$:
\begin{eqnarray}
\hat P_<(2n-1,s)&=&{\tau_-^{-1}}\hat Q_+\hat Q_- [\hat P_+\hat
P_-]^{n-1},\label{podd}\\
\hat P_<(2n,s)&=&{\tau_-^{-1}}\hat Q_-^2P_+ [\hat P_+\hat
P_-]^{n-1},\label{peven}
\end{eqnarray}
where
\begin{equation}
\hat Q_\pm(s)=\frac{1-\hat P_\pm(s)}{s},
\end{equation}
is the Laplace transform of Eq.~(\ref{defq}). If we express the
conservation of probability,
\begin{equation}
P_<(t)+\sum_{N=1}^{+\infty} P_<(N,t)=1,\label{norma}
\end{equation}
and after summing simple geometric series, we obtain,
\begin{equation}
\hat P_<(s)=\frac{1}{s}-\frac{1-\hat P_-(s)}{\tau_- \,s^2}.
\label{pinf}
\end{equation}
This relation is nothing but the Laplace transform of
Eq.~(\ref{pnew2}). It is certainly satisfying, and also reassuring,
that the IIA reproduces this exact relation, as well as the
equivalent relation between $P_>(t)$ and $P_+(t)$ of
Eq.~(\ref{pnew1}). Of course, $P_>(2n-1,t)$ and $P_>(2n,t)$ satisfy
similar equations as
Eqs.~(\ref{conv1},\ref{conv2},\ref{podd},\ref{peven},\ref{norma},\ref{pinf}),
obtained by exchanging the indices $+ \leftrightarrow -$ and $<\,
\leftrightarrow\, >$.

Using Eqs.~(\ref{conda1},\ref{conda2},\ref{condn1},\ref{condn2}), we
can now write explicitly the Laplace transform of the known
quantities $A_<(t)$ and $N_<(t)$ in terms of the Laplace transform
of $P_\pm(t)$:
\begin{eqnarray}
\hat N_<(s)=\frac{(1+\hat P_+)(1-\hat P_-)}{\tau_- \,s^2(1-\hat
P_+\hat P_-)},\label{eq0}\\
\hat A_<(s)=G(M)\left[\frac{1}{s}-\frac{1-\hat P_+}{1+\hat P_+}\,
\hat N_<(s) \right].\label{eq1}
\end{eqnarray}
Again, $A_>(t)$ and $N_>(t)$ are given by similar expressions, after
the substitution $+ \leftrightarrow -$ and $<\, \leftrightarrow\,
>$, and $G(M) \leftrightarrow \bar G(M)$.

Using $\hat P_\pm'(0)=-\int_0^{+\infty}t\,P_\pm(t)\,dt=-\tau_\pm$
and Eq.~(\ref{t2}), one obtains the following estimates when $s\to
0$,
\begin{equation}
\hat N_<(s)\sim\frac{1}{\tau \,s^2},\quad\hat
A_<(s)\sim\frac{G^2(M)}{s}.\label{est}
\end{equation}
The first expression in Eq.~(\ref{est}) is equivalent to the general
result of Eq.~(\ref{as}), whereas the second relation expresses that
for large $t$, $A_<(t)\sim G^2(M)$. For large $s$,
\begin{equation}
\hat N_<(s)\sim \frac{1}{\tau_- s^2},\quad
\hat A_<(s)\sim\frac{G(M)}{s},\label{estl}
\end{equation}
which corresponds to the small time behavior of Eq.~(\ref{smalln})
for $N_<(t)$, whereas the second relation is equivalent to
$A_<(0)=G(M)$.

Finally, writing
\begin{equation}
\hat F_<(s)=\frac{G(M)-s\,\hat A_<(s)}{G(M)\,s\,\hat N_<(s)},\label{eq2}
\end{equation}
and using Eqs.~(\ref{eq0},\ref{eq1}), the interval distributions are
given by
\begin{eqnarray}
\hat P_+(s)&=&\frac{1-\hat F_<(s)}{1+ \hat F_<(s)}, \label{pp} \\
\hat P_-(s)&=&\frac{2-\tau_- \,s^2 \hat N_<(s)(1+\hat F_<(s))}
{2-\tau_-\,s^2 \hat N_<(s)(1-\hat F_<(s))}.\label{pm}
\end{eqnarray}
Inserting these expressions of $\hat P_\pm(s)$ in Eq.~(\ref{pinf}),
one obtains  $\hat P_<(s)$ (and $\hat P_>(s)$), from the sole
knowledge of $A_<(t)$ and $N_<(t)$ (or their Laplace transform),
which are known explicitly for a Gaussian process. Alternative
expressions for $\hat P_\pm$ in terms of the Laplace transform of
$A_>(t)$ and $N_>(t)$ are readily obtained after the substitution $+
\leftrightarrow -$ and $<\,\leftrightarrow\, >$, and $G(M)
\leftrightarrow \bar G(M)$ in
Eqs.~(\ref{est},\ref{estl},\ref{eq2},\ref{pp},\ref{pm}). Finally,
due to the symmetry of the process under the transformation $X\to
-X$, the following symmetric relations hold:
\begin{eqnarray}
P_\pm(-M,t)=P_\mp(M,t),\quad P_>(-M,t)=P_<(M,t).\label{symM}
\end{eqnarray}

Using Eq.~(\ref{dapf}), we find that the dimensionless function
$W(t)$,
\begin{eqnarray}
W(t)&=&-\tau\,A'_>(t)=-\tau\,A'_<(t),\nonumber\\
&=& -\frac{\tau}{2\pi}\frac{f'(t)}{\sqrt{1-f^2(t)}}
\exp\left(-\frac{M^2}{1+f(t)}\right), \label{defW}\\
&=&\frac{a}{2}\,
{\rm e}^{-\frac{M^2}{2}.\frac{1-f}{1+f}},\label{W}
\end{eqnarray}
has a simpler analytical form than $A_>(t)$ or $A_<(t)$. Similarly,
we define the dimensionless auxiliary function, $V(t)$, by
\begin{eqnarray}
\tau\,\bar G(M)N'_>(t)&=&\tau\,\langle |X'(t)| \rangle_{>}=
\bar G(M)+V(t),\label{defV1}\\
\tau\,G(M)N'_<(t)&=&\tau\,\langle |X'(t)| \rangle_{<}=
G(M)-V(t).\label{defV2}
\end{eqnarray}
Using Eq.~(\ref{Xm}), we find that $V(t)$ takes an explicit form
contrary to $N_>(t)$ or $N_<(t)$:
\begin{eqnarray}
V(t)=G(M)-G(b)+a
\left[G\left(a\,b\right)-\frac 12\right]
{\rm e}^{-\frac{M^2}{2}.\frac{1-f}{1+f}},\label{V}
\end{eqnarray}
where $a(t)$ and $b(t)$ are simple functionals of the correlator
$f(t)$, which have been defined in Eqs.~(\ref{defa},\ref{defb}).
We give below the behavior of $W(t)$ and $V(t)$, in the limit $t\to +\infty$,
\begin{eqnarray}
V(t)&\sim &\frac{M}{\sqrt{2\pi}}\,{\rm e}^{-\frac{M^2}{2}}f(t),\label{asV}\\
W(t)&\sim &-\frac{1}{2\sqrt{a_2}}\,{\rm e}^{-\frac{M^2}{2}} f'(t),\label{asW}
\end{eqnarray}
while one has $V(0)=G(M)-1/2$ and $W(0)=1/2$.

The Laplace transform of $W(t)$ and $V(t)$ can be explicitly written
\begin{eqnarray}
\tau^{-1}\,\hat W(s)=G(M)-s\hat A_<(s)=\bar G(M)-s\hat A_>(s),
\end{eqnarray}
and
\begin{eqnarray}
\tau\,\bar G(M)\,s\,\hat N_>(s)&=&\frac{\bar G(M)}{s}+\hat V(s),\\
\tau\,G(M)\,s\,\hat N_<(s)&=&\frac{G(M)}{s}-\hat V(s).
\end{eqnarray}
In terms of $\hat W(s)$ and $\hat V(s)$, the interval distributions
take the symmetric form
\begin{eqnarray}
\hat P_+(s)&=&\frac{G(M)-s\,\hat V(s)-s\,\hat W(s)}
{G(M)-s\,\hat V(s)+s\,\hat W(s)}, \label{pp1} \\
\hat P_-(s)&=&\frac{\bar G(M)+s\,\hat V(s)-s\,\hat W(s)}
{\bar G(M)+s\,\hat V(s)+s\,\hat W(s)}.\label{pm1}
\end{eqnarray}
In practice, the explicit forms of $W(t)$ and $V(t)$ obtained in
Eqs.~(\ref{W},\ref{V}) permit a fast and efficient numerical
implementation of the IIA.

Note that for a smooth \emph{non-Gaussian} process, \emph{all the
above results of the IIA remain unaltered}, $G(M)$ now being the
cumulative sum of the associated distribution of $X$, and $\tau$
being given by the general form \cite{csp},
\begin{equation}
\tau^{-1}=g(M)\langle |X'(t)|\rangle,\label{taugen}
\end{equation}
whereas $\tau_\pm$ are still given by Eq.~(\ref{t2}). Applying the
IIA results to a general non-Gaussian process only requires the
prior knowledge of $A_<(t)$ and $N_<(t)$ (or $A_>(t)$ and $N_>(t)$).
In general, these time-dependent functions should be given \emph{a
priori}, analytically, or extracted from numerical or experimental
data.

Finally, let us briefly address the validity of the IIA. First,
crossing intervals are strictly \emph{never} independent, except in
the particular case of a Markovian process (see Eq.~(\ref{markov})),
for which the IIA does not apply, due to the singular nature of this
process. The IIA is also an uncontrolled approximation which seems
almost impossible to improve systematically, by introducing interval
correlations. However, in practice, the IIA is found numerically to
be a surprisingly good approximation, especially for ``very smooth''
processes for which $f(t)$ is analytic \cite{iia,csp} (see the
counterexample of $f_3(t)$ in Eq.~(\ref{f3}) below). We will even
show in the next sections that some of the predictions of the IIA
are in fact exact for smooth Gaussian processes.

\section{Persistence exponents}\label{III}

\subsection{General properties and physical applications}

In many contexts, one is interested in the large time behavior of
the persistence probabilities $P_<(t)$ and $P_>(t)$. It has been
rigorously established that, if $|f(t)|\leq C/t$, for sufficiently
large time $t$ ($C$ is some arbitrary constant), then the
persistence decays exponentially \cite{slepian}. Hence, we define
the two persistence exponents, by the asymptotics
\begin{equation}
P_<(t)\sim{\rm e}^{-\theta_- t}, \quad P_>(t)\sim{\rm e}^{-\theta_+ t},
\label{thdef}
\end{equation}
valid when $t\to+\infty$. Due to the symmetry relation of
Eq.~(\ref{symM}), we have
\begin{eqnarray}
\theta_\pm(-M)=\theta_\mp(M).\label{symth}
\end{eqnarray}
Hence, from now, we only consider the case $M\geq 0$. From
Eqs.~(\ref{pnew1},\ref{pnew2},\ref{pnew3},\ref{pnew4}), we find that
the interval distributions $P_\pm(t)$ decays in the same way as
their associated persistence, for $t\to+\infty$.

The name persistence ``exponent'' (instead of ``decay rate'') arises
from its numerous applications in out of equilibrium physics
\cite{SM,AB1,BD1,per1,per2,per3,iia,globalc,global,red2,
krug,csp,red1,breath,lq,soap,diff1d,surf}.
Indeed, in many cases, the normalized two-times correlation function
of the relevant physical variable $Z(T)$ obeys \emph{dynamical
scaling},
\begin{equation}
\frac{\langle Z(T)\,Z(T')\rangle}{\sqrt{\langle Z^2(T)\rangle
\langle Z^2(T)\rangle}}=F(T/T'),
\label{scaling}
\end{equation}
where $T$ is the physical time, and $F$ is the scaling correlation
function. Defining
\begin{equation}
t=\ln T, \quad X(t)=\frac{Z(T)}{\sqrt{\langle Z^2(T)\rangle}},
\end{equation}
the resulting process $X(t)$ becomes \textit{stationary} in the new effective
time $t$ \cite{per1,per2}, with correlator
\begin{equation}
f(t)=F(\exp t).
\end{equation}
Hence, the persistence $P_>^X(M,t)$ for the process $X(t)$ is equal
to the probability $P_>^Z(M,T)$ that the process $Z(T)$ remains
above the level $M{\times}\sqrt{\langle Z^2(T)\rangle}$ up to time
$T=\exp t$ \cite{per1,red2}. Since $P_>^X(M,t)$ decays exponentially, the persistence
of the process $Z(T)$ decays as a power-law, hence the name
persistence ``exponent'',
\begin{equation}
P_>^Z(M,T)=P_>^X(M,t)\sim {\rm e}^{-\theta_+ t}\sim T^{-\theta_+ }.
\end{equation}
Similarly, one has
\begin{equation}
P_<^Z(M,T)=P_<^X(M,t)\sim {\rm e}^{-\theta_- t}\sim T^{-\theta_- }.
\end{equation}
In particular, for $M=0$, the persistence of the processes $X$ and
$Z$ are both equal to the probability that the associated process
does not change sign up to the time $t=\ln T$.

In order to illustrate the dynamical scaling of the correlator
resulting from Eq.~(\ref{scaling}), let us give three physical
examples. In the next sections, our analytical results will be
tested on the correlators $f_1(t)$, $f_2(t)$, and $f_3(t)$,
introduced below in Eqs.~(\ref{f1},\ref{f2},\ref{f3}).
\begin{itemize}
\item[$\bullet$] Consider a $d$-dimensional ferromagnetic system
(for instance, modelized by the Ising model) quenched from its
equilibrium state above the critical temperature ${\cal T}_c$,
down to ${\cal T}_c$ (critical quench) or below ${\cal T}_c$
(subcritical quench). As time $T$ proceeds, correlated domains
of linear size $L(T)\sim T^{1/z}$ grow, and this coarsening
dynamics leads to dynamical scaling for the total magnetization,
of the form Eq.~(\ref{scaling}). Initially the spins have only
short range spatial correlations, and as the domains grow, the
correlation length remains finite, of order $L(T)$. If $L(T)$
remains much smaller than the linear size of the system, the law
of large numbers ensures that the magnetization $Z(T)$,
which is the sum of the individual spins, is a true
Gaussian variable. For $M=0$, the persistence is equivalent to
the probability that the magnetization never changes sign from
the time of the quench ($T=0$), up to time $T$. For a critical
quench, the persistence decays as a power-law with a persistence
exponent $\theta_c$, which is a universal critical exponent of
spin systems, independent of the familiar ones ($\beta$, $\eta$,
$z$,...), due to the non-Markovian nature of the magnetization
\cite{globalc,per3}. For a subcritical quench, the magnetization
persistence also decays as a power-law with a universal
$d$-dependent persistence exponent controlled by a zero
temperature fixed point \cite{global}, and the dynamical
exponent is $z=2$.

\item[$\bullet$] If the field $Z({\bf x},T)$ evolves according to
the $d$-dimensional diffusion equation (or more sophisticated
interface model equations \cite{krug}),
\begin{equation}
\frac{\partial Z}{\partial T}=\nabla_{\bf x}^2 Z,
\end{equation}
starting from an initial random configuration of zero mean, the
process becomes Gaussian for large times, another consequence of
the law of large numbers. For any fixed ${\bf x}$, the
normalized two-times correlator of $Z({\bf x},T)$ obeys
dynamical scaling, and the probability that $Z({\bf x},T)$ does
not change sign decays as a power-law with a $d$-dependent
exponent computed approximately in \cite{iia}. The associated
stationary correlator in the variable $t=\ln T$ is
\begin{eqnarray}
f_1(t)=\frac{1}{\cosh^{\frac d 2}\left(\frac{t}{2}\right)}.\label{f1}
\end{eqnarray}
Moreover, in $d=1$ and at a fixed time $T$, the process $Z(x,T)$
is a stationary Gaussian process in the spatial variable $x$.
The variable $X(x)=Z(x,T)/\sqrt{\langle Z^2(x,T)\rangle_x}$
(where $\langle \,.\,\rangle_x$ denotes the average over the
spatial variable $x$) has a Gaussian correlator. Hence, we shall
later illustrate our results using the correlator
\begin{eqnarray}
f_2(t)={\rm e}^{-\frac{t^2}{2}}.\label{f2}
\end{eqnarray}

\item[$\bullet$] The random acceleration process \cite{theta2}
is defined by its equation of motion,
\begin{equation}
\frac{d^2Z}{dT^2}=\eta(T),\label{randomacc}
\end{equation}
where $\eta(T)$ is a $\delta$-correlated white noise. Again, its
two-times correlator obeys dynamical scaling, and the associated
stationary correlator is \cite{globalc},
\begin{eqnarray}
f_3(t)=\frac{3}{2}\,{\rm e}^{-\frac{|t|}{2}}-\frac{1}{2}\,{\rm
e}^{-\frac{3|t|}{2}}.\label{f3}
\end{eqnarray}
For $M=0$, this process is a rare case for which the exact value
of the persistence exponent is known exactly \cite{theta2},
\begin{equation}
\theta_\pm(M=0)=\frac 1 4.\label{sinai}
\end{equation}
Note that for the random acceleration process, the correlator
$f_3(t)$ is not analytic. Although, twice differentiable at
$t=0$, its third derivative is not defined at $t=0$. For this
process, it is not surprising to find that the IIA is not as
precise as for smoother processes \cite{iia,csp}. Indeed, one
finds the IIA result $\theta_\pm(M=0)=0.2647...$, off by 6\,\%
compared to Eq.~(\ref{sinai}), a relative error much bigger than
usually observed for $M=0$ persistence exponents obtained by
means of the IIA.
\end{itemize}

\subsection{Persistence exponents within the IIA}

Within the IIA, the persistence exponents $\theta_\pm$ are obtained
as the first pole on the negative real axis of the Laplace transform
of the associated interval distribution $P_\pm(t)$, since the
Laplace transform of $\exp(-\theta_\pm t)$ is $1/(s+\theta_\pm)$.
Using Eqs.~(\ref{pp},\ref{pm}), we find that $\theta_\pm$ satisfies
\begin{eqnarray}
G(M)[1+\theta_- \hat N_<(-\theta_-)]+ \theta_-
\hat A_<(-\theta_-)&=&\frac{1}{\tau\,\theta_-}, \label{thetam1}\\
G(M)[1-\theta_+ \hat N_<(-\theta_+)]+ \theta_+
\hat A_<(-\theta_+)&=&0, \label{thetap1}
\end{eqnarray}
or the equivalent relations in terms of $\hat A_>$ and $\hat N_>$
\begin{eqnarray}
\bar G(M)[1+\theta_+ \hat N_>(-\theta_+)]+ \theta_+
\hat A_>(-\theta_+)&=&\frac{1}{\tau\,\theta_+}, \label{thetap2}\\
\bar G(M)[1-\theta_- \hat N_>(-\theta_-)]+ \theta_-
\hat A_>(-\theta_-)&=&0.\label{thetam2}
\end{eqnarray}
In terms of the auxiliary functions $V(t)$ and $W(t)$ introduced in
Eqs.~(\ref{defW},\ref{defV1},\ref{defV2}), the defining equations of
$\theta_\pm$ take a simpler form
\begin{eqnarray}
\hat W(-\theta_+)-\hat V(-\theta_+)&=&\frac{G(M)}{\theta_+}, \label{thp} \\
\hat W(-\theta_-)+\hat V(-\theta_-)&=&\frac{\bar G(M)}{\theta_-}.\label{thm}
\end{eqnarray}
The residues $R_\pm$ associated to $\theta_\pm$, and defined by
\begin{eqnarray}
P_\pm(t)\sim R_\pm\, {\rm e}^{-\theta_\pm t},
\end{eqnarray}
can easily be extracted from Eqs.~(\ref{pp1},\ref{pm1}), using the
identity
\begin{eqnarray}
R_\pm^{-1}=\frac{\,d\hat P_\pm^{-1}}{ds}(s=-\theta_\pm).
\end{eqnarray}

Before addressing the limit $M\to\pm\infty$ in the next section, we
present some numerical results for moderate $M$, and for the
processes associated to the correlators $f_1(t)$, $f_2(t)$, and
$f_3(t)$, introduced in Eqs.~(\ref{f1},\ref{f2},\ref{f3}). We recall
the symmetry relation $\theta_\pm(-M)=\theta_\mp(M)$, so that we
restrain ourselves to the case $M\geq 0$. In order to compare the
values of $\theta_\pm$ for the different correlators, it is
instructive to multiply the persistence exponents (of dimension
$[t]^{-1}$) by the time scale $\tau_0$, which is the mean crossing
time interval for $M=0$,
\begin{equation}
\tau_0=\tau(M=0)=\frac{\pi}{\sqrt{a_2}}.\label{tau0}
\end{equation}

We have performed extensive numerical simulations of the processes
associated to the correlators $f_1(t)$, $f_2(t)$, and $f_3(t)$, and
measured the persistence and the crossing intervals distributions,
and in particular, the persistence exponents. Let us briefly describe
how to generate long trajectories of a stationary Gaussian process
solely characterized by its two-times correlator \cite{per1}. In
real time, the most general form of such a process $X$ reads
\begin{equation}
X(t)=\int_{-\infty}^tJ(t-t')\,\eta(t')\,dt',\label{kernel}
\end{equation}
where $\eta(t)$ is a $\delta$-correlated Gaussian white noise. The
Gaussian nature of $\eta(t)$ and the linear form of
Eq.~(\ref{kernel}) ensures that $X(t)$ is a Gaussian process.
Moreover, the convolution product of the noise $\eta(t')$ with the
kernel $J(t-t')$ (instead of a general kernel $J(t,t')$) ensures
stationarity. Taking the Fourier transform of Eq.~(\ref{kernel}), we
obtain
\begin{equation}
\tilde{X}(\omega)=\tilde J(\omega)\,\tilde\eta(\omega),\label{fourier}
\end{equation}
where $\tilde X(\omega)=\int_{-\infty}^{+\infty}X(t)\exp(-i\omega
t)\,dt$, and the noise Fourier transform satisfies $\langle
\tilde\eta(\omega)\,\tilde\eta(\omega')
\rangle=2\pi\,\delta(\omega+\omega')$. The Fourier transform of the
correlator of $X$ is hence
\begin{equation}
\frac{\langle  \tilde X(\omega)\,\tilde X(\omega') \rangle}{2\pi} =\tilde  f(\omega)\,
\delta(\omega+\omega')=|\tilde  J(\omega)|^2 \,\delta(\omega+\omega'),
\end{equation}
which relates the kernel $J(t)$ to the correlation function $f(t)$,
through their Fourier transform, $\tilde  f(\omega)=|\tilde
J(\omega)|^2$. Note that a necessary condition for $f(t)$ to be the
correlator of a Gaussian process is that its Fourier transform $\tilde
f(\omega)$ remains positive for all real frequencies $\omega$. Finally, a
trajectory of $X$ is obtained after sampling
\begin{equation}
\tilde X(\omega)=\sqrt{\tilde  f(\omega)}\,\tilde \eta(\omega),\label{fourier1}
\end{equation}
on a frequency mesh, and performing an inverse fast Fourier
transform of the obtained $\tilde  X(\omega)$. The Fourier transform of
the correlators $f_1(t)$, $f_2(t)$, and $f_3(t)$ having simple
explicit expressions, this procedure for obtaining long trajectories
is extremely efficient. Otherwise, one has to tabulate the Fourier
transform of $f(t)$, before simulating Eq.~(\ref{fourier1}).

In practice, our numerical results are obtained after averaging
$10^6$--$10^7$ trajectories of length ${\cal T}=1024\,\tau_0$
(sometimes $2048\,\tau_0$) and with a frequency mesh of typically
$1024^2$ points spaced by $\Delta\omega={\cal T}^{-1}$. Since for a
general $M$, $\tau=\tau_0\exp(M^2/2)$, we obtain typically
$10^9$--$10^{10}{\times}\exp(-M^2/2)$ $M$-crossings, a number which decays
rapidly for large $M$. Despite the loss of statistics for large
positive $M$, we find that $\theta_-(M)$ can still be measured with
great accuracy, since $P_-(t)$ and $P_<(t)$ becomes purely
Poissonian in this limit (see next section). On the other hand, the
error bars for $\theta_+(M)$ increase rapidly with $M$, due to the
occurrence of less $M$-crossings. In addition, the determination of
$\theta_+(M)$ is also plagued by the fact that the exponential
asymptotics of $P_+(t)$ and $P_>(t)$ only develops for increasingly
large $t$, as $M$ increases, which may produce uncontrolled
systematic errors in the numerical estimates of $\theta_+(M)$ (see next
section).

Table~\ref{tab1} compares $\tau_0\,\theta_-(M)$ for the correlators
$f_1(t)$ (for $d=2$) and $f_2(t)$, as obtained from numerical
simulations and from the IIA calculation of Eq.~(\ref{thm}). The
agreement between the theory and the simulations is excellent for
both correlators, and is even improving as $M$ increases. In fact,
we will show in the next section that the IIA becomes exact for
$\theta_-(M)$, when $M\to\infty$, and we will obtain an analytic
asymptotic expression for $\theta_-(M)$. Our theoretical and
numerical results are also consistent with the numerical bounds
computed in \cite{merca} for the correlator $f_1(t)$, for $M=1$ and
$M=2$:
\begin{eqnarray}
0.3681<\tau_0\,\theta_{-,1}(M=1)<0.4298,\\
0.0666<\tau_0\,\theta_{-,1}(M=2)<0.0748.
\end{eqnarray}
It is also clear that the IIA provides much better estimates of the
persistence exponent $\theta_-(M)$ than these bounds, which are
however exact, although they require a much bigger numerical effort
than the IIA \cite{merca}.

Table~\ref{tab2} compares $\tau_0\,\theta_+(M)$ for the correlators
$f_1(t)$ and $f_2(t)$, as obtained from numerical simulations and
from the IIA calculation of Eq.~(\ref{thp}). The agreement between
the theory and the simulations is satisfying, although we will show
that the IIA ultimately fails in predicting the exact asymptotics of
$\theta_+(M)$ as $M\to\infty$, although the limiting form of
$P_+(t)$ will be given exactly by the IIA, for $t$ not too large.

\begin{table}
\begin{tabular}{cccccc}
$M$& $\tau_0\,\theta_{-,1}^{{\rm IIA}}$ & $\tau_0\,\theta_{-,1}^{{\,\rm sim}}$ &
$\tau_0\,\theta_{-,2}^{{\rm IIA}}$ & $\tau_0\,\theta_{-,2}^{{\,\rm sim}}$\\
\hline 0 & $1.1700$ & 1.178(2) &
1.2928 & $1.330(5)$ \\
1/2 & 0.6949 & 0.7008(7)
  & 0.7587 & 0.7723(8) \\
1 & $0.3715$ & $0.3743(6)$ &
$0.3994$ & $ 0.4032(8)$ \\
3/2 & 0.1734 & 0.1750(4)
  & 0.1831 &  0.1850(4) \\
2 & $6.813\cdot 10^{-2}$ & $6.834(6)\cdot 10^{-2}$ &
$7.065\cdot 10^{-2}$ & $7.116(7)\cdot 10^{-2}$ \\
5/2 & $2.177\cdot 10^{-2}$ & $2.180(3)\cdot 10^{-2}$
  & $2.224\cdot 10^{-2}$ & $2.225(3)\cdot 10^{-2}$ \\
3 & $5.509\cdot 10^{-3}$ & $5.510(2)\cdot 10^{-3}$ &
$5.568\cdot 10^{-3}$ & $5.568(2)\cdot 10^{-3}$ \\
\end{tabular}
\caption{\label{tab1} We report the value of $\tau_0\,\theta_-(M)$
as obtained from the IIA calculation ($\theta_{-}^{{\rm IIA}}$) and
simulations ($\theta_{-}^{{\,\rm sim}}$), for different values of
$M$, calculated for the processes associated to the correlators
$f_1(t)$ ($\tau_0=2\pi$) and $f_2(t)$ ($\tau_0=\pi$), introduced in
Eqs.~(\ref{f1},\ref{f2}).}
\end{table}

\begin{table}
\begin{tabular}{cccccc}
$M$& $\tau_0\,\theta_{+,1}^{{\rm IIA}}$ & $\tau_0\,\theta_{+,1}^{{\,\rm sim}}$ &
$\tau_0\,\theta_{+,2}^{{\rm IIA}}$ & $\tau_0\,\theta_{+,2}^{{\,\rm sim}}$\\
\hline 0 & $1.1700$ & 1.178(2) &
1.2928 & $1.330(5)$ \\
1/2 & 1.8164 &  1.865(6)
  & 2.0232  &  2.125(7) \\
1 & 2.6475 &   2.67(2)
  & 2.9606 & 3.11(2) \\
3/2 & 3.6677 & 3.74(3)
  & 4.1031 &  4.36(4) \\
2 &  4.8926 &  5.06(5)
  & 5.4338 &  5.90(7) \\
5/2 &  6.2651 & 6.6(1)
  & 6.9152 &  7.6(2) \\
\end{tabular}
\caption{\label{tab2} We report the value of $\tau_0\,\theta_+(M)$
as obtained from the IIA calculation ($\theta_{+}^{{\rm IIA}}$) and
simulations ($\theta_{+}^{{\,\rm sim}}$), for different values of
$M$, calculated for the processes associated to the correlators
$f_1(t)$ ($\tau_0=2\pi$) and $f_2(t)$ ($\tau_0=\pi$), introduced in
Eqs.~(\ref{f1},\ref{f2}).}
\end{table}

For the process associated to the correlator $f_3(t)$
($\tau_0=2\pi/\sqrt 3$), $\tau_0\,\theta_\pm^{{\rm
IIA}}(M=0)=0.9602...$, compared to the exact value
$\tau_0\,\theta_\pm(M=0)=\frac{\pi}{2\sqrt{3}}=0.9069...$ obtained in
\cite{theta2}. The agreement between the IIA and numerical estimates
of $\theta_-(M)$ greatly improves as $M$ increases, as observed for
the two other correlators in Table~\ref{tab1}, whereas $\theta_+(M)$
is only fairly reproduced for large $M$ (as already observed in
Table~\ref{tab2}), suggesting that the IIA somewhat fails to
reproduces $\theta_+(M)$ in the limit $M\to+\infty$, as will be
confirmed in the next section. Finally, we note that for the three
processes considered here (and all other smooth Gaussian processes
known to the author), one has $\theta_\pm(M=0)\sim \tau_0^{-1}$,
with a proportionality constant close to unity.

\subsection{Exact results in the limit $M\to\pm\infty$}

In the limit $M\to+\infty$, and using
\begin{equation}
\bar G(M)\sim\frac{{\rm e}^{-\frac{M^2}{2}}}{\sqrt{2\pi}\, M},
\end{equation}
and Eqs.~(\ref{taugauss},\ref{t2}), the mean of the $\pm$ intervals
are
\begin{eqnarray}
\tau_-&\sim &\frac{2\pi}{\sqrt{a_2}}\,{\rm e}^{\frac{M^2}{2}}\sim 2\tau,
\label{taumlarge}\\
\tau_+&\sim &\sqrt{\frac{2\pi}{a_2}}\,M^{-1}.
\end{eqnarray}
Hence, as expected physically, the typical length of the $-$
intervals is becoming increasingly large as $M\to+\infty$ (of order
$2\tau$), whereas the typical length of the $+$ is going slowly to
zero.

\subsubsection{Distribution of $-$ intervals and $\theta_-$}

For large $M$, and using Eqs.~(\ref{W},\ref{V}), we obtain that
\begin{equation}
V(t)=W(t)\left(1+O\left({\rm e}^{-\frac{M^2}{2}}\right)\right)\label{largeM}.
\end{equation}
Moreover, in this limit both functions can be approximated by
developing $f(t)$ up to second order in $t$,
\begin{equation}
V(t)\sim W(t)\sim\frac{1}{2}\,{\rm e}^{-\frac{M^2a_2t^2}{8}}=
\frac{1}{2}\,{\rm
e}^{-\frac{\pi}{4}\left(\frac{t}{\tau_+}\right)^2}.\label{Wap}
\end{equation}
The constraint that
\begin{eqnarray}
\tau^{-1}\int_0^{+\infty} W(t)\,dt&=&A_<(0)-A_<(+\infty),\\
&=&A_>(0)-A_>(+\infty),\\
&=&G(M)\bar G(M),\\
&\sim &\frac{{\rm e}^{-\frac{M^2}{2}}}{\sqrt{2\pi}\, M},
\end{eqnarray}
is consistently recovered up to leading order, after integrating
Eq.~(\ref{Wap}). Moreover, the explicit form of Eq.~(\ref{Wap})
implies that the Laplace transform of $W(t)$ can be approximated by
its value at $s=0$, provided that
\begin{equation}
|s|\ll \tau_+^{-1},\quad M\gg 1.\label{condvalm}
\end{equation}
Finally, using  Eqs.~(\ref{t2},\ref{pm1}), we find that the Laplace
transform of $P_-(t)$ is given by
\begin{equation}
\hat P_-(s)=\frac{1}{1+s\,\tau_-},\label{poisson}
\end{equation}
under the conditions of Eq.~(\ref{condvalm}). Hence, we conclude
that for large $M$, one has
\begin{equation}
\theta_-(M)\sim \frac{1}{\tau_-}\sim\frac{1}{2\tau},\label{asthm}
\end{equation}
and, taking the inverse Laplace transform of Eq.~(\ref{poisson}),
that the distribution of $-$ intervals is essentially Poissonian
\begin{equation}
P_-(t)=\tau_-^{-1}{\rm e}^{-\frac{t}{\tau_-}},\quad P_<(t)={\rm
e}^{-\frac{t}{\tau_-}},\label{poissonfin}
\end{equation}
except in a narrow region of time $0\leq t\leq \tau_+\ll\tau_-$,
corresponding to the conjugate time domain  of the condition of
Eq.~(\ref{condvalm}). The behavior of $P_-(t)$ and $P_<(t)$, for
$t\ll\tau_+$, will be obtained exactly in section~\ref{secdist}. The
fact that the distribution of the long $-$ intervals is becoming
Poissonian can be physically interpreted. Indeed, since
$\tau_-\to+\infty$ when $M\to+\infty$, the process $X$ can be
considered to be Markovian at this time scale. In addition, this
also shows that for the $-$ intervals, the IIA is in fact exact. In
the next section, in the process of finding exact bounds for
$\theta_+(M)$, we will prove the exact result, valid in the opposite
$M\to 0$ limit,
\begin{equation}
\theta_-(M)=\theta_0- \frac{\langle X(t)\rangle_{X>0}}{2\hat f(0)}\,M
+O\left(M^2\right),\label{exactexpm1}
\end{equation}
where $\theta_0=\theta_-(M=0)$, $\langle X(t)\rangle_{X>0}$ is
the average of $X$ over all trajectories for which $X(t)>0$ for all
times, and $\hat f(0)=\int_0^{+\infty}f(t)\,dt$.

\begin{figure}
\psfig{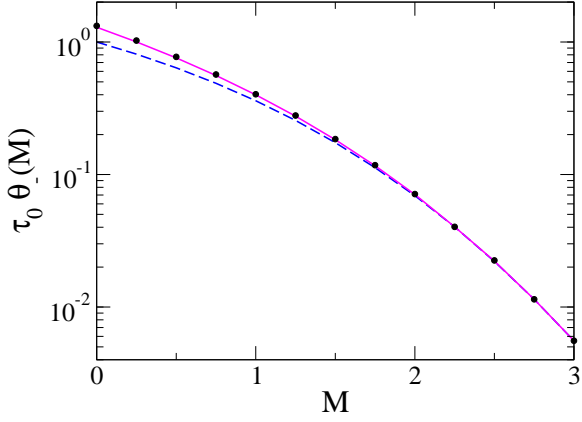} \caption{\label{fig1} (Color online)
We plot $\theta_-(M)$ (in unit of $\tau_0^{-1}$), for the non-Markovian
process associated to the correlator $f_2(t)$ of Eq.~(\ref{f2}).
Symbols are the results of numerical simulations (see text and
Eq.~(\ref{fourier1})), while the full line is the result of the IIA
approximation, Eq.~(\ref{thm}). Finally, the dashed line corresponds
to the exact asymptotic result, $\theta_-(M)\sim \tau_-^{-1}$. Some
values of $\theta_-(M)$ are also reported on Table~\ref{tab1}.}
\end{figure}

\begin{figure}
\psfig{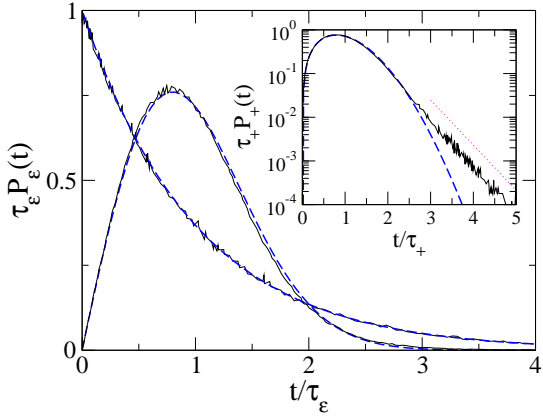} \caption{\label{fig2} (Color online) For
$M=3$, we plot the distribution of $\varepsilon=\pm$ intervals,
$P_\pm(t)$, as a function of $t/\tau_\pm$ (full lines; for $M=3$,
$\tau_+\approx 0.76350\, a_2^{-1/2}$ and $\tau_-\approx 564.83\,
a_2^{-1/2}$), for the process associated to the correlator $f_2(t)$
of Eq.~(\ref{f2}). The distribution of $-$ intervals has converged
to the Poissonian form of Eq.~(\ref{poissonfin}) (dashed line). Note
that the linear regime of $P_-(t)$ predicted in
Eqs.~(\ref{res1},\ref{res3}), for $t\ll\tau_+\ll\tau_-$, cannot be seen at
this scale. The distribution of $+$ intervals is well described by
the Wigner distribution of Eq.~(\ref{wigner}) (dashed line), but
ultimately decays exponentially for $t\gg a_2^{-1/2}$ (insert;
dotted line of slope $\tau_+\,\theta_+\approx 2.35$).}
\end{figure}

On Fig.~\ref{fig1}, we plot $\theta_-(M)$ obtained by simulating the
process associated to the correlator $f_2(t)$ of Eq.~(\ref{f2}). We
also plot the IIA result of Eq.~(\ref{thm}), which is in perfect
agreement with numerical simulations. In addition, we illustrate the
rapid convergence of $\theta_-(M)$ to the exact asymptotics of
Eq.~(\ref{asthm}). On Fig.~\ref{fig2}, we plot $P_-(t)$ for $M=3$,
which has already perfectly converged to its asymptotic Poissonian
form of Eq.~(\ref{poissonfin}).

\subsubsection{Distribution of $+$ intervals and $\theta_+$}

For $M\to+\infty$, and using the asymptotics Eq.~(\ref{largeM}) and
the IIA expression for $\hat P_+(s)$ of Eq.~(\ref{pp1}), we obtain,
\begin{equation}
\hat P_+(s)=1-2\,s\,\hat W(s).\label{pps1}
\end{equation}
In real time, Eq.~(\ref{pps1}) reads
\begin{equation}
P_+(t)=-2\tau A''_>(t)=\frac{\pi}{2}\frac{t}{\tau_+^2}
{\rm e}^{-\frac{\pi}{4}\left(\frac{t}{\tau_+}\right)^2},\label{wigner}
\end{equation}
which is valid for $t\ll a_2^{-1/2}$. Using
Eqs.~(\ref{ap},\ref{taumlarge}), the relation $P_+(t)=-2\tau
A''_>(t)$ can be explicitly written as
\begin{equation}
P_+(t)=-\tau_-\langle X'(t)X'(0)\,\delta(X(t)-M)\, \delta
(X(0)-M)\rangle. \label{ppapprox}
\end{equation}
When the process $X$ crosses the level $M$ for the first time at
time $t$, after the preceding crossing at time 0, one has
$-X'(t)X'(0)>0$. Eq.~(\ref{ppapprox}) states that for small times
$t\ll a_2^{-1/2}$, the probability of having a $+$ interval of
length $t$ is of the same order as the probability of having an
$M$-crossing at time $t$, knowing that there is such a crossing at
time $0$, which follows a $-$ interval (hence the factor $\tau_-$).
In other words, for short time $t$, a crossing at time $t$ is very
often the first crossing following the crossing at time $0$.

The form of $P_+(t)$ found in Eq.~(\ref{wigner}) can be obtained
from a simple physical argument, which will be our basis for
obtaining exact results for the small $t$ behavior of $P_\pm(t)$ in
section~\ref{secdist}. Indeed, let us consider a $+$ interval of
length  $t\ll a_2^{-1/2}$. For $0\leq t'\leq t$,
one can expand the \textit{smooth} process
$X(t')$ in power of $t'$ up to second order, starting from $X(0)=M$,
\begin{equation}
X(t')=M+X'(0)\,t'+\frac{X''(0)}{2}\,t'^2+...\label{trajec}
\end{equation}
remembering that $X'(t)$ is a Gaussian variable of second moment
$\langle X'^2(t)\rangle=a_2=-f''(0)$, the probability distribution
of the velocity $v=X'(0)>0$, at a $M$-crossing following a $-$
interval, is given by
\begin{equation}
\rho(v)=\frac{\left\langle |X'(0)|\delta(X(0)-M)\,
\delta(X'(0)-v)\right\rangle_{X'(0)>0}} {\left\langle
|X'(0)|\delta(X(0)-M)\right\rangle_{X'(0)>0}}.
\end{equation}
Since $X(0)$ and $X'(0)$ are uncorrelated (as $\langle
X'(0)X(0)\rangle=f'(0)=0$), these Gaussian averages are easily
performed, leading to the exact result
\begin{equation}
\rho(v)=\frac{v}{a_2}\,{\rm e}^{-\frac{v^2}{2a_2}}.\label{distv}
\end{equation}
In addition, the distribution of $X''(0)$, conditioned to the fact
that $X(0)=M$, is a Gaussian of mean
\begin{equation}
\langle X''(0) \rangle_{X(0)=M}=-Ma_2,
\end{equation}
and mean square deviation $f^{IV}(0)-f''^2(0)$, which is of order
$a_2^2$, and is independent of $M$. Hence, for large $M$, one can
replace $X''(0)$ by its average, and the interval length $t$ can be
obtained by finding the first $M$-crossing of the trajectory of
Eq.~(\ref{trajec}),
\begin{equation}
t_v=\frac{2v}{Ma_2}.
\end{equation}
Hence, for small time $t$, on has
\begin{equation}
P_+(t)=\int_0^{+\infty}\rho(v)\,\delta(t-t_v)\,dv.
\end{equation}
Finally, using Eq.~(\ref{distv}) and the asymptotic expression for
$\tau_+$ of Eq.~(\ref{taumlarge}), we obtain the distribution of $+$
intervals given by Eq.~(\ref{wigner}).

Thus we find that the distribution of $+$ intervals is given by the
Wigner distribution for $t\ll a_2^{-1/2}$, a result also obtained in
\cite{rice}. However, note that the ratio $t/\tau_+$ can be
arbitrarily large in the limit $M\to +\infty$, or $\tau_+\to 0$. The
probability distribution of Eq.~(\ref{wigner}) is correctly
normalized to unity and has a mean equal to $\tau_+$. Of course, for
$t\gg a_2^{-1/2}$, the actual distribution of $+$ interval should
decay exponentially \cite{slepian}, with a rate $\theta_+(M)$.
Matching the two asymptotics at $t\sim a_2^{-1/2}$, we find that up
to a so far unknown multiplicative constant,
\begin{equation}
\theta_+(M)\sim \sqrt{a_2}\,M^2,\label{asthetap}
\end{equation}
for large $M$. In addition, the above argument shows that the total
probability contained in the exponential tail of $P_+(t)$ vanishes
extremely rapidly as $M\to+\infty$, and is of order $\exp(
-K\,M^2)$, where $K$ is a constant of order unity. Moreover, the
result of Eq.~(\ref{wigner}) implies that the persistence is given
by
\begin{equation}
P_>(t)=
{\rm e}^{-\frac{\pi}{4}\left(\frac{t}{\tau_+}\right)^2},\label{wigner2}
\end{equation}
for $t\ll a_2^{-1/2}$, and decays exponentially for $t\gg
a_2^{-1/2}$.

In the limit $M\to+\infty$, the determination of $\theta_+(M)$ seems
to be beyond the IIA. Indeed, let us assume for simplicity that
$f(t)$ decays faster than any exponential, $f(t)\sim
\exp(-c\,t^\gamma)$, with $\gamma>1$ . Anticipating that
$\theta_+(M)$ is large, we need to evaluate $\hat P_+(s)$ for large
negative $s$. In this limit, Eqs.~(\ref{asV},\ref{asW}) leads to
\begin{eqnarray}
\hat V(s)&\sim &\frac{M}{\sqrt{2\pi}}\,{\rm e}^{-\frac{M^2}{2}}\hat f(s),\\
\hat W(s)&\sim &-\frac{1}{2\sqrt{a_2}}\,{\rm e}^{-\frac{M^2}{2}}s\,\hat f(s).
\end{eqnarray}
If $1<\gamma<2$, and using the IIA expression for $\hat P_+(s)$ of
Eq.~(\ref{pp1}), we obtain
\begin{equation}
\hat P_+(s)\sim \frac{2\,\sqrt{\frac{2\,a_2}{\pi}}\,M}
{s+\sqrt{\frac{2\,a_2}{\pi}}\,M},\label{poissonp}
\end{equation}
which leads to
\begin{equation}
\theta_+^{\rm IIA}(M)\sim \sqrt{\frac{2\,a_2}{\pi}}\,M,
\end{equation}
which grossly underestimate the divergence of $\theta_+(M)$, when
$M\to\infty$. If $f(t)\sim \exp(-c\,t^2)$ decays as a Gaussian ($\gamma=2$),
Eq.~(\ref{pp1}) leads to
\begin{equation}
\theta_+^{\rm IIA}(M)\sim \sqrt{c}\,M,\label{iiagaussc}
\end{equation}
which, again, behaves linearly with $M$. Finally, if $\gamma>2$, we find that
\begin{equation}
\theta_+^{\rm IIA}(M)\sim \,M^{2\frac{\gamma-1}{\gamma}},
\end{equation}
up to a computable multiplicative constant.

Let us now present exact bounds for $P_>(t)$ which will lead to an
exact asymptotics for $\theta_+(M)$, fully consistent with
Eq.~(\ref{asthetap}). We discretize time $t_i=i\Delta t$, with
$\Delta t= t/n$, and define $x_i=X(t_i)$. By definition,
\begin{equation}
P_>(t,M)=\int_M^{+\infty}\frac{\prod_{i=1}^n dx_i}{(2\pi)^{n/2}\sqrt{\det {\bf C}}}
\,{\rm e}^{-S(\{x_i\})},
\end{equation}
where the Gaussian ``action'' has the quadratic form
\begin{equation}
S(\{x_i\})=\frac 12\sum_{i,j}D_{ij}\,x_i x_j,
\end{equation}
and where the matrix ${\bf D}$ is the inverse of the correlation
matrix ${\bf C}$ defined by its matrix elements
\begin{equation}
C_{ij}=\langle X(t_i)X(t_j)\rangle=f(t_i-t_j).
\end{equation}
Making the change of variables $y_i=x_i+M\in[0,+\infty]$, and noting
that
\begin{equation}
S(\{y_i\})=\frac 12\sum_{i,j}D_{ij\,}y_i y_j+M\sum_{i,j}D_{ij}\,y_i
+\frac{M^2}{2}\sum_{i,j}D_{ij},
\end{equation}
we obtain
\begin{equation}
P_>(t,M)=P_>(t,M=0)
\left\langle{\rm e}^{-M \sum_i \sigma_i y_i}\right\rangle_{y>0}\,
{\rm e}^{-\frac{M^2}{2}n\bar\sigma},\label{exact1}
\end{equation}
with
\begin{equation}
\sigma_i=\sum_{j=1}^n D_{ij},\quad \bar\sigma=\frac{1}{n}\sum_{i=1}^n \sigma_i,
\end{equation}
and where $\langle \,.\,\rangle_{y>0}$ denotes the average over all
processes for which $y(t_i)\geq 0$, for all $i$. If we assume that
the process is periodic of period $t$, the vector ${\bf
u}=(1,1,...,1)$ is an exact eigenvector of ${\bf C}$ associated to
the eigenvalue
\begin{equation}
\lambda=\sum_{i=-n/2}^{n/2} f(t_i),\label{eigen}
\end{equation}
and one has
\begin{equation}
\sigma_i=\sum_{j=1}^n D_{ij}{\times} 1={\bf (D.u)}_i=\lambda^{-1}{u}_i=\lambda^{-1}.
\end{equation}
For large time $t$, the periodic constraint should not affect the
value of the persistence exponent. In fact, in this limit of large
time and fine discretization ($\Delta t\to 0$), and even dropping
the assumption that the process is periodic, one finds that
\begin{equation}
\sigma_i=\bar\sigma=\lambda^{-1}=
\frac{\Delta t}{\int_{-\infty}^{\infty} f(t)\,dt}=
\frac{\Delta t}{2\hat f(0)},
\end{equation}
where the discrete sum of Eq.~(\ref{eigen}) has been transformed
into an integral when $\Delta t\to 0$, and the integral limits $\pm
t/2$ extended to $\pm \infty$ for large $t$. Finally, defining
$\theta_0=\theta_\pm(M=0)$ and using Eq.~(\ref{exact1}), we find the
exact result
\begin{equation}
\theta_+(M)=\theta_0+\Theta(M)+\frac{M^2}{4\hat f(0)},\label{exact4}
\end{equation}
with
\begin{equation}
\Theta(M)=\lim_{t\to+\infty}-\frac 1 t\,\ln
\left\langle{\rm e}^{-\frac{M}{2\hat f(0)}
\int_0^t X(t')\,dt'}\right\rangle_{X>0}.
\label{exact2}
\end{equation}
Using the convexity of the exponential function, and the fact that
the argument in the exponential in Eq.~(\ref{exact2}) is negative,
we obtain the exact bounds
\begin{equation}
0\leq \Theta(M) \leq
\frac{\langle X(t)\rangle_{X>0} }{2\hat f(0)}\,M,
\label{exact3}
\end{equation}
where $\langle X(t)\rangle_{X>0}$ is the average of the process $X$
restricted to the trajectories which remain positive for all times, and
is a constant strictly independent of $M$.
Eqs.~(\ref{exact4},\ref{exact3}) lead to an exact bound for
$\theta_+(M)$,
\begin{equation}
\theta_0+\frac{M^2}{4\hat f(0)}\leq\theta_+(M)
\leq\theta_0+\frac{\langle X(t)\rangle_{X>0} }{2\hat f(0)}\,M
+\frac{M^2}{4\hat f(0)},\label{exactbound}
\end{equation}
which implies that for large $M$,
\begin{equation}
\theta_+(M)\sim\frac{M^2}{4\hat f(0)},\label{exact5}
\end{equation}
with a subleading correction bounded by a linear term in $M$. The
exact asymptotics of Eq.~(\ref{exact5}) confirms our heuristic
argument of Eq.~(\ref{asthetap}). In addition, for small $M$, it is
clear from Eqs.~(\ref{exact2},\ref{exact3}) that one has the exact
expansion,
\begin{equation}
\theta_+(M)=\theta_0+\frac{\langle X(t)\rangle_{X>0} }{2\hat f(0)}\,M
+O\left(M^2\right).\label{exactexp}
\end{equation}
Since $\theta_-(M)=\theta_+(-M)$, we also get
\begin{equation}
\theta_-(M)=\theta_0-\frac{\langle X(t)\rangle_{X>0} }{2\hat f(0)}\,M
+O\left(M^2\right).\label{exactexpm}
\end{equation}

\begin{figure}
\psfig{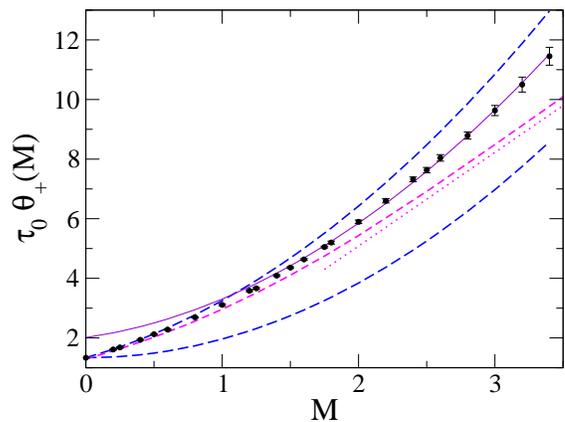} \caption{\label{fig3} (Color online) We
plot $\theta_+(M)$ (in unit of $\tau_0^{-1}$), for the non-Markovian
process associated to the correlator $f_2(t)$ of Eq.~(\ref{f2}).
Symbols correspond to results of numerical simulations. The upper
and lower dashed line are the exact bounds of Eq.~(\ref{exactbound})
(the upper bound being exact up to order $M$, for small $M$). The
ratio of these two bounds goes to $1$ when $M\to+\infty$. The full
line is a quadratic fit for $M>3/2$, to the functional form
$\theta_+(M)=a_0+a_1\,M +\frac{M^2}{4\hat f(0)}$, with $a_0\approx
2.0$ and $a_1\approx 0.66$. Finally, the middle dashed line is the
IIA result (along with its asymptotic slope given by
Eq.~(\ref{iiagaussc}); dotted line), which underestimates the
quadratic growth of $\theta_+(M)$. Some values of $\theta_+(M)$ are
also reported on Table~\ref{tab2}.}
\end{figure}

On Fig.~\ref{fig3}, we plot $\theta_+(M)$ obtained by simulating the
process associated to the correlator $f_2(t)$ of Eq.~(\ref{f2}). We
also plot the IIA result of Eq.~(\ref{thm}), which underestimates
the actual value of $\theta_+(M)$, as explained above. In addition,
we plot the exact bounds of Eq.~(\ref{exactbound}), as well as a
convincing fit of $\theta_+(M)$, to the functional form
$\theta_+(M)=a_0+a_1\,M +\frac{M^2}{4\hat f(0)}$, where the exact
leading term was obtained in Eq.~(\ref{exact5}). On Fig.~\ref{fig2},
we plot $P_+(t)$ for $M=3$, which follows the predicted Wigner
distribution of Eq.~(\ref{wigner}), for $t\ll a_2^{-1/2}$, before
decaying exponentially, with rate $\theta_+(M)$, for large $t$.

\section{Distributions $P_\pm(t)$ for small intervals}\label{secdist}

The heuristic argument presented in the preceding section, below
Eq.~(\ref{trajec}), can be adapted to provide the exact behavior of
$P_\pm(t)$, $P_>(t)$, and $P_<(t)$ for small time $t$. For a smooth
process with $f(t)$ at least four times differentiable, we will show
that the IIA surprisingly reproduces these exact results. However,
for a marginally smooth process, such that for small $t$, and
$2<\alpha< 4$,
\begin{equation}
f(t)=1-\frac{a_2}{2}\,t^2+a_\alpha\,|t|^\alpha+...\label{singcusp}
\end{equation}
the fourth derivative of $f(t)$ is not defined at $0$. We will show
that for such a process, the IIA does not lead to exact results for
the small $t$ behavior of $P_\pm(t)$, although it is in fact
qualitatively and even quantitatively accurate. Note that the
process associated to the correlator $f_3(t)$ defined in
Eq.~(\ref{f3}) satisfies the property of Eq.~(\ref{singcusp}), with
$\alpha=3$, which implies that its velocity is not differentiable.

\subsection{Exact small time behavior for very smooth processes}

In the limit $t\to 0$, the trajectory of the very smooth process $X$
inside a $+$ or $-$ interval is essentially parabolic,
\begin{equation}
X(t')=M+v\,t'+\frac{a}{2}\,t'^2+O\left(t^3\right),\label{trajecex}
\end{equation}
where the distribution of the velocity at a crossing time is given
by
\begin{equation}
\rho(v)=\frac{|v|}{a_2}\,{\rm e}^
{-\frac{v^2}{2a_2}},
\end{equation}
and the distribution of the acceleration $a$, conditional to the
fact that $X(0)=M$, is
\begin{equation}
\sigma(a)=\frac{1}{\sqrt{2\pi}\,a_2 z}\,{\rm e}^
{-\frac{\left(a+Ma_2\right)^2}{2a_2^2z^2}},
\end{equation}
with
\begin{equation}
z=\sqrt{\frac{f^{IV}(0)}{f''^2(0)}-1},\label{zdef}
\end{equation}
and $a_2=-f''(0)$. Note that the acceleration at $t'=0$ is
independent of the velocity at $t'=0$, since $\langle
X''(0)X'(0)\rangle=f'''(0)=0$. In addition, since
\begin{equation}
f^{IV}(0)-f''^2(0)=\int_{-\infty}^{+\infty}
\left(\omega^2-a_2\right)^2\tilde f(\omega)\,\frac{d\omega}{2\pi}\,>0,
\end{equation}
$z$ defined by Eq.~(\ref{zdef}) is indeed a positive real number.

Let us first consider the small time behavior of $P_+(t)$. For an
interval of length $t$ to be small, and since $v>0$, the
acceleration is necessarily negative. From Eq.~(\ref{trajecex}), the
crossing time is then given by
\begin{equation}
t=-\frac{2v}{a},
\end{equation}
which is only valid when $t$ is small. Hence, for small $t$,
\begin{equation}
P_+(t)=\int_{0}^{+\infty}dv\int_{-\infty}^{0}da\,\,\rho(v)\,\sigma(a)
\,\delta\left(t+\frac{2v}{a}\right).
\end{equation}
After integrating over $v$, we obtain
\begin{equation}
P_+(t)=\frac{1}{2}\int_{-\infty}^{0}
\rho\left(\frac{a\,t}{2}\right)\sigma(a)|a|\,da
\end{equation}
Using the explicit form of $\rho(v)$, and taking the limit $t\to 0$,
we obtain
\begin{equation}
P_+(t)=c_+(M)\,a_2\,t+O\left(t^3\right),\label{res1}
\end{equation}
where the dimensionless constant $c_+(M)$ is given by
\begin{equation}
c_+(M)=\frac{1}{4\,a_2^2}\int_{-\infty}^{0}\sigma(a)a^2\,da.
\end{equation}
Performing explicitly the Gaussian integral above, we finally obtain
the exact result,
\begin{equation}
c_+(M)=\frac{M^2+z^2}{4}\,
G\left(\frac Mz\right)+\frac{z\,M}{4} \,g\left(\frac Mz\right),\label{res2}
\end{equation}
where $g(X)$ is the Gaussian distribution and $G(X)$ its cumulative
sum, both defined in Eqs.~(\ref{gauss1},\ref{gauss2}). In the limit
$M\to+\infty$, we find $c_+(M)\sim M^2/4$, which leads to
\begin{equation}
P_+(t)\sim \frac{M^2\,a_2}{4}\,t\sim\frac{\pi}{2}\frac{t}{\tau_+^2},
\end{equation}
in agreement with our result of Eq.~(\ref{wigner}). Moreover, using
Eq.~(\ref{pnew1}), we obtain the small time expansion of $P_>(t)$,
up to third order in time,
\begin{equation}
P_>(t)=1-\frac{t}{\tau_+}+\frac{c_+\,a_2}{6\,\tau_+}\,t^3+O\left(t^5\right).
\end{equation}
Finally, the corresponding results for $c_-(M)$, $P_-(t)$, and
$P_<(t)$ are obtained by the substitution $M\leftrightarrow -M$, and
the exchange of the indices $+\leftrightarrow -$ and
$>\,\leftrightarrow \,<$. In particular, we have
\begin{equation}
c_-(M)=\frac{M^2+z^2}{4}\,
\bar G\left(\frac Mz\right)-\frac{z\,M}{4} \,g\left(\frac Mz\right).\label{res3}
\end{equation}
In the limit $M\to+\infty$, we find that
\begin{equation}
c_-(M)\sim\frac{z^5}{2M^3} \,g\left(\frac Mz\right).
\end{equation}

Thus, we find that $P_\pm(t)$ behaves linearly with time for small
$t$, a result which will be shown in section~\ref{marginal} to be
specific to very smooth processes, for which the correlator $f(t)$
is at least four times differentiable at $t=0$.

\begin{figure}
\psfig{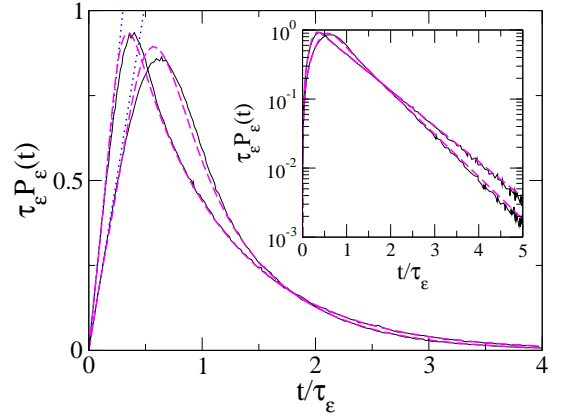} \caption{\label{fig4} (Color online) For
$M=1/2$, we plot the distribution of $\varepsilon=\pm$ intervals,
$P_\pm(t)$, as a function of $t/\tau_\pm$ (full lines), for the
process associated to the correlator $f_2(t)$ of Eq.~(\ref{f2})
($P_-(t)$ is the most peaked distribution). The straight dotted
lines have the predicted slopes at $t=0$, given by
Eqs.~(\ref{res1},\ref{res2},\ref{res3}). The dashed lines are the
distributions obtained from the IIA, after taking the inverse Laplace
transform of Eqs.~(\ref{pp1},\ref{pm1}). The insert shows the same
data on a semi-log scale, illustrating the good accuracy of the IIA
in predicting the persistence exponents $\theta_\pm(M)$ and their
associated residue $R_\pm$.}
\end{figure}

On Fig.~\ref{fig4}, for $M=1/2$, we plot $P_\pm(t)$ obtained from
numerical simulations of the process associated to the correlator
$f_2(t)$, illustrating their linear behavior for small time $t$. The
initial slope at $t=0$ is in perfect numerical agreement with the
exact results of Eqs.~(\ref{res1},\ref{res2},\ref{res3}). In
addition,  we also plot the full distributions $P_\pm(t)$, obtained
by taking the inverse Laplace transform of
Eqs.~(\ref{pp1},\ref{pm1}). For this moderate value of $M$, far from
the large $M$ regime where the IIA becomes exact, the good agreement
between the IIA results and numerical simulations is certainly
encouraging.

\subsection{IIA results}

We now derive the small time behavior of $P_\pm(t)$ by using the
IIA. Expanding the explicit form of $V(t)$ an $W(t)$ of
Eqs.~(\ref{W},\ref{V}), we find,
\begin{eqnarray}
W(t)&=&  \frac{1}{2}-\frac{M^2+z^2}{16}\,a_2\,t^2+O\left(t^4\right)  ,\\
V(t)&=&  G(M)-\frac{1}{2} -\left[\frac{M^2+z^2}{8}\,
\left(G\left(\frac Mz\right)-\frac 12\right)\right.\nonumber\\
&&+\left.\frac{z\,M}{8} \,g\left(\frac Mz\right)\right]\,a_2\,t^2+O\left(t^4\right).
\end{eqnarray}
Taking the Laplace transform of these expression and inserting them
into the IIA expression for $\hat P_+(s)$ of Eq.~(\ref{pp1}), we
obtain, for large $s$,
\begin{eqnarray}
\hat P_+(s)=\left[\left(M^2+z^2\right)\,
G\left(\frac Mz\right)\right.\nonumber\\
\left. + \,z\,M\,g\left(\frac Mz\right)\right]\,\frac{a_2}{4\,s^2}+
O\left(s^{-4}\right),
\end{eqnarray}
which is exactly the Laplace transform of
Eqs.~(\ref{res1},\ref{res2}). Thus, we find that the IIA reproduces
the exact small time behavior of $P_\pm(t)$.

\subsection{Marginally smooth processes}\label{marginal}

In this section, we study marginally smooth processes characterized
by a correlator $f(t)$ having a small time expansion of the form of
Eq.~(\ref{singcusp}), so that $f^{IV}(0)$ does not exist. Since
$f^{IV}(0)$ was explicitly appearing in our results of the preceding
sections, this suggests that the small time behavior of $P_\pm(t)$
should be affected by the weak singularity in $f(t)$.

Let us first apply the IIA in this marginal case. The small time
expansion of $V(t)$ and $W(t)$ now read,
\begin{eqnarray}
W(t)&\sim &  \frac{1}{2}-\frac{(\alpha-1)a_\alpha}{2\,a_2}\,t^{\alpha-2} ,\\
V(t)& \sim &  G(M)-\frac{1}{2}
-M\sqrt{\frac{(\alpha-1)a_\alpha}{4\pi}}\,t^{\alpha/2}.
\end{eqnarray}
Note that since  $2<\alpha<4$, one has $\alpha-2<\alpha/2$. For
large $s$, and using Eq.~(\ref{pp1}) and the above asymptotics, we
obtain,
\begin{eqnarray}
\hat P_\pm(s)\sim \frac{(\alpha-1)\Gamma(\alpha-1)a_\alpha}{2\,a_2\,s^{\alpha-2}}.
\end{eqnarray}
In real time, this leads to the small time behavior,
\begin{eqnarray}
P_\pm(t)\sim \frac{(\alpha-1)\Gamma(\alpha-1)a_\alpha}{2\,\Gamma(\alpha-2)\,a_2}
\,t^{\alpha-3},\label{iiaalpha}
\end{eqnarray}
which is independent of $M$. In particular, for the quite common
case $\alpha=3$, which corresponds to the correlator $f_3(t)$
introduced in Eq.~(\ref{f3}), we find that $P_\pm(t)$ should be
constant at $t=0$, with
\begin{eqnarray}
P_\pm(0)=\frac{a_3}{a_2}.\label{iiaalpha1}
\end{eqnarray}
For the correlator $f_3(t)$, one has $a_2=3/4$ and $a_3=1/4$, so
that $P_\pm(0)=1/3$.

Let us now derive an exact expression for $P_+(t)$ for a marginally
smooth process with $\alpha=3$. For small $t$, the correlator of the
velocity is
\begin{equation}
\langle X'(t)X'(0)\rangle=-f''(t)=a_2-6a_3|t|+O\left(t^{2}\right),\label{cormar}
\end{equation}
which coincides with the small time behavior of the correlator of a
Markovian process (see Eq.~(\ref{markovcor})). Hence, on short time
periods, the local equation of motion of $X'(t)$ is,
\begin{equation}
X''(t)=-\frac{6\,a_3}{a_2}\,X'(t)+2\sqrt{3\,a_3}\,\eta(t),\label{motion}
\end{equation}
where $\eta(t)$ is a $\delta$-correlated Gaussian noise. From the
equation of motion  Eq.~(\ref{motion}), one indeed recovers
Eq.~(\ref{cormar}), for short time. Now using Eq.~(\ref{motion}), a
short time trajectory of $X$, starting from $X(0)=M$ and $X'(0)=v$,
takes the form,
\begin{equation}
X(t)=M+v\,t+2\sqrt{3\,a_3}\,Z(t)+O\left(t^{2}\right),
\end{equation}
where $Z(t)=O\left(t^{3/2}\right)$ is the random acceleration
process introduced in
Eq.~(\ref{randomacc}),
\begin{equation}
Z(t)=\int_0^t dt_1\int_0^{t_1} dt_2\,\eta(t_2).
\end{equation}
Finally, for small $t$, the first $M$-crossing of the process $X$,
corresponds to the first time for which $Z(t)/t$ crosses the level
\begin{equation}
Z_0=-\frac{v}{2\sqrt{3\,a_3}}.
\end{equation}
Introducing the probability distribution $\Psi(t_0,Z_0)$ that
$Z(t)/t$ crosses $Z_0$ for the first time at time $t=t_0$, one has
the scaling relation
\begin{equation}
\Psi(t_0,Z_0)=\frac{1}{Z_0^2}\psi\left(\frac{t_0}{Z_0^2}\right),
\end{equation}
obtained by noticing that the scale invariant process $Z(t)$ has
dimension $[t]^{3/2}$. For small time $t$,
\begin{equation}
P_\pm(t)=\int_0^{+\infty}
\rho(v)\,\psi\left(\frac{12\,a_3\, t}{v^2}\right)\frac{12\,a_3}{v^2}\, dv.
\end{equation}
After making the change of variable $T=12\,a_3\, t/v^2$ and taking
the limit $t\to 0$, while using the fact that $\rho(v)\sim v/a_2$
for small $v$, we obtain the final exact result,
\begin{equation}
P_\pm(0)=\frac{6\,a_3}{a_2}\int_0^{+\infty}
\psi\left(T\right)\,\frac{dT}{T}=\frac{6\,a_3}{a_2}\,
\left\langle \frac 1T\right\rangle.
\end{equation}
Up to a dimensional constant $6\,a_3/a_2$ depending on $f(t)$,
$P_\pm(0)$ is proportional to the mean inverse first-passage time of
the process $Z(t)/t$ at the level $Z_0=1$. By simulating the process
$Z$, we have obtained
\begin{equation}
\left\langle \frac 1T\right\rangle\approx 0.193(1),\label{iianumt}
\end{equation}
which leads to
\begin{equation}
P_\pm(0)=1.158(6)\frac{a_3}{a_2}.\label{iianumt2}
\end{equation}
This result has also been checked numerically for the process
associated to $f_3(t)$. In fact, the constant appearing in
Eq.~(\ref{iianumt2})  was obtained exactly by Wong \cite{wong},
based on the study of the process $Z(t)$ of \cite{kean}. Their
result leads to
\begin{equation}
P_\pm(0)=\frac{37}{32}\,\frac{a_3}{a_2}=1.15625\, \frac{a_3}{a_2}.\label{iianumt3}
\end{equation}

Hence, we find that the IIA result of
Eq.~(\ref{iiaalpha1}) is not exact for marginally smooth processes
with $\alpha=3$, although it predicts correctly that $P_\pm(0)$ is a
constant independent of $M$, leading to a reasonably accurate
estimate of this constant. For general $\alpha$,
Eq.~(\ref{iiaalpha}) is certainly correct dimensionally speaking,
and probably fairly accurate, in practice.

We end this section by an approximate calculation of $\left\langle
1/T\right\rangle$ for the process $Z(t)$, which does not reproduce
the exact result of \cite{wong}, obtained by a much more complex
method \cite{BL,kean}. We make the approximation,
\begin{equation}
Z(t)=\int_0^t dt_1\int_0^{t_1} dt_2\,\eta(t_2)\approx \frac{t}{\sqrt 3}
 \int_0^t dt_1\, \eta(t_1),
\end{equation}
where the factor $1/\sqrt 3$ ensures that both processes have the
same mean square displacement $\langle Z^2(t)\rangle=t^3/3$. Then,
the original first-passage problem for $Z(t)/t$ becomes a standard
first-passage problem for the usual Brownian motion $B(t)$ at the
level $B_0= \sqrt 3$, for which the first-passage time probability
distribution is given by \cite{red1},
\begin{equation}
\Psi\left(T, B_0\right)=\frac{B_0}{\sqrt{2\pi}\,T^{3/2}}\,
{\rm e}^{-\frac{B_0^2}{2T}},
\end{equation}
 and for which,
\begin{equation}
\left\langle \frac 1T\right\rangle=\frac{1}{B_0^2}.
\end{equation}
Finally, within this simple approximation, we find that
$\left\langle 1/T\right\rangle=1/3$, overestimating the value
obtained in Eq.~(\ref{iianumt}).

\section{Conclusion}

In this work, we have considered the $M$-crossing interval
distributions and the persistence of a smooth non-Markovian Gaussian
process. We have obtained exact results for the persistence
exponents in the limit of a large crossing level $M$, including
exact bounds for $\theta_+(M)$ and $P_>(t)$. In this limit, we have
shown that the distributions of $+$ and $-$ intervals become
universal, and are respectively given by the Wigner and Poisson
distributions. For any value of $M$, we have obtained the exact
small time behavior of the interval distributions and the
persistence. We have also derived these results within the
Independent Interval Approximation. Quite surprisingly, the IIA
reproduces all these exact results, except for the large $M$
asymptotics of $\theta_+(M)$. In addition, the IIA fails in
reproducing the exact small time behavior of the interval
distributions for marginally smooth processes, although it remains
qualitatively correct and even quite accurate in this case. To the
credit of the IIA, it is the only method to provide precise
approximate expressions of the interval distributions and the
persistence for all times, and all values of the level $M$, and thus
to grant access to the distribution of extrema of a non-Markovian
Gaussian process. In addition, the IIA can be straightforwardly
applied to any \emph{smooth non-Gaussian process}, for which the
autocorrelation function $A_>(t)$ (or $A_<(t)$) and the conditional
number of crossings $N_>(t)$ (or $N_<(t)$) are known analytically,
or extracted from experimental or numerical data. For Gaussian
processes, simple forms of the derivative of these quantities have
been obtained, which permits a simple and fast numerical
implementation of the IIA results.

\acknowledgments I am very grateful to Satya Majumdar and Partha
Mitra for fruitful discussions.


\begin{thebibliography}{0}

\bibitem{adler} R.~J. Adler, {\it The Geometry of Random Fields} (John
Wiley \& Sons Ltd., Chichester, 1981); R.~J. Adler and J.~E. Taylor,
{\it Random Fields and Geo\-me\-try} (Springer, Berlin, 2007).


\bibitem{BL} I.~F. Blake and W.~C. Lindsey, IEEE
Trans. Info. Theory {\bf 19}, 295 (1973); S.~O. Rice, {\it Noise
and Stochastic Processes}, p.~133, Ed. by N. Wax (Dover, NY,
1954); C.~W. Helstrom, {IRE Trans. Info. Theory} {\bf IT-3}, 232
(1957); J.~A. McFadden, IRE Trans. Info. Theory {\bf IT-4}, 14
(1957).

\bibitem{rice} S.~O. Rice, Bell Syst. Tech. J. \textbf{37}, 581 (1958).

\bibitem{kean} H.~P. McKean,  J. Math. Kyoto  Univ. {\bf 2}, 227 (1963).

\bibitem{wong} E. Wong, SIAM J. Appl. Math. \textbf{14}, 1246 (1966).

\bibitem{slepian} D. Slepian, Bell Syst. Tech. J. {\bf 41}, 463 (1962).

\bibitem{theta2} T.~W. Burkhardt, {J. Phys. A} {\bf 26}, L1157 (1993);
Y.~G. Sinai, {Theor. Math. Phys.} {\bf 90}, 219 (1992).

\bibitem{merca} C. Mercadier, Adv. in Appl. Probab. {\bf 38},
149 (2006).

\bibitem{SM} S.~N. Majumdar, Current Science {\bf 77}, 370 (1999).

\bibitem{AB1} A.~J. Bray, B. Derrida, and C. Godr\`eche, {J. Phys. A} {\bf
27}, L357 (1994).

\bibitem{BD1} B. Derrida, V. Hakim, and V. Pasquier, {Phys. Rev. Lett.}
{\bf 75}, 751 (1995).

\bibitem{per1} S.~N. Majumdar and C. Sire,  Phys. Rev. Lett. {\bf 77},
1420 (1996).

\bibitem{per2} C. Sire, S.~N. Majumdar, and A. R\"udinger,
Phys.  Rev. E {\bf 61}, 1258 (2000).

\bibitem{per3} K. Oerding, S.~J. Cornell, and A.~J. Bray, {Phys. Rev. E}
{\bf 56}, R25 (1997).

\bibitem{iia}  S.~N. Majumdar, C. Sire, A.~J. Bray, and  S.~J. Cornell,
Phys. Rev. Lett. {\bf 77}, 2867 (1996); B. Derrida, V. Hakim, and
R. Zeitak, Phys. Rev. Lett. {\bf 77}, 2871 (1996).

\bibitem{globalc} S.~N. Majumdar, A.~J. Bray,  S.~J. Cornell, and C. Sire,
{Phys. Rev. Lett.} {\bf 77}, 3704 (1996)

\bibitem{global} S. Cueille and C. Sire, {J. Phys. A}  {\bf 30}, L791 (1997).

\bibitem{red2} P.~L. Krapivsky and S. Redner, Am. J. Phys. \textbf{64}, 546
(1996).

\bibitem{krug}  J. Krug, H. Kallabis, S.~N. Majumdar, S.~J. Cornell, A.~J.
Bray, and C. Sire, {Phys. Rev. E} {\bf 56}, 2702 (1997); S.~N.
Majumdar and A.~J. Bray, Phys. Rev. Lett. {\bf 86}, 3700 (2001);
S.~N. Majumdar and C. Dasgupta, Phys. Rev. E \textbf{73}, 011602 (2006).

\bibitem{csp} C. Sire, Phys. Rev. Lett. {\bf 98}, 020601 (2007).

\bibitem{red1} S. Redner, \emph{A Guide to First-Passage Processes}
(Cambridge University Press, Cambridge, 2001).

\bibitem{breath} M. Marcos-Martin, D. Beysens, J.-P. Bouchaud, C.
Godr\`eche, and I. Yekutieli, {Physica A} {\bf 214}, 396 (1995).

\bibitem{lq}  B. Yurke, A.~N. Pargellis, S.~N. Majumdar, and C. Sire,
{Phys. Rev. E} {\bf 56}, R40 (1997).

\bibitem{diff1d} G.~P. Wong, R.~W. Mair, R.~L. Walsworth, and
D.~G. Cory, Phys. Rev. Lett. {\bf 86}, 4156 (2001).

\bibitem{surf} D.~B. Dougherty, I. Lyubinetsky, E.~D. Williams,
M. Constantin, C. Dasgupta, and S. Das Sarma, Phys. Rev. Lett. {\bf 89},
136102 (2002).

\bibitem{soap} W.~Y. Tam, R. Zeitak, K.~Y. Szeto, and J. Stavans,
{Phys. Rev. Lett.} {\bf 78}, 1588 (1997).


\bibitem{max1} K.~L. Chung, Bull. Amer. Math. Soc. {\bf 81}, 742 (1975);
K.~L. Chung, Ark. Mat. {\bf 14}, 155 (1976);  D.~P. Kennedy, J.
Appl. Probab. \textbf{13}, 371 (1976).

\bibitem{max2} S.~N. Majumdar, J. Randon-Furling,
M.~J. Kearney, and Marc Yor, preprint \emph{arXiv:0802.2619} (2008).

\bibitem{Gumbel} E.~J. Gumbel, {\em Statistics of Extremes} (Columbia
University Press, New York, 1958).

\bibitem{Galambos} J. Galambos, {\em The Asymptotic Theory of Extreme
Or\-der Statistics} (R.~E. Krieger Publishing Co., Malabar, 1987).

\bibitem{extreme} A. Comtet, P. Leboeuf, and S.~N. Majumdar, Phys.
Rev. Lett. \textbf{98}, 070404 (2007); D.-S. Lee, Phys. Rev. Lett.
\textbf{95}, 150601 (2005); C.~J. Bolech and A. Rosso, Phys. Rev.
Lett. \textbf{93}, 125701 (2004); D.~S. Dean and S.~N. Majumdar,
Phys. Rev. E \textbf{64}, 046121 (2001); S.~N. Majumdar and P.~L.
Krapivsky, Phys. Rev. E {\bf 62}, 7735 (2000).

\bibitem{image} K.~J. Worsley, Ann. Statist. {\bf 23}, 640 (1995);
Adv. Appl. Prob. {\bf 27}, 943, (1995).

\bibitem{lander} E.~S. Lander and D. Botstein, Genetics {\bf 121},
185 (1989).

\bibitem{boupot} J.-P. Bouchaud and M. Potters, \emph{Theory
of Financial Risk and Derivative Pricing: From Statistical Physics
to Risk Management} (Cambridge University Press, Cambridge, 2003).


\bibitem{potts} C. Sire and  S.~N. Majumdar, Phys. Rev. Lett. {\bf
74}, 4321 (1995); Phys. Rev. E {\bf 52}, 244 (1995).




\end{thebibliography}
\end{document}